\documentclass[lettersize,journal]{IEEEtran}
\usepackage{microtype}
\usepackage{graphicx}
\usepackage{subfigure}
\usepackage{booktabs} 
\usepackage{algorithm}
\usepackage{amsmath}
\usepackage{algpseudocode}
\usepackage{algorithm}
\usepackage{algorithmicx}
\usepackage{algpseudocode}
\usepackage{verbatim}
\usepackage{graphicx}
\usepackage{multirow}
\usepackage{amsfonts}
\usepackage{cite}
\usepackage{color}
\usepackage{microtype}
\usepackage{graphicx}
\usepackage{subfigure}
\usepackage{booktabs} 
\usepackage{algorithm}
\usepackage{amsmath}
\usepackage{algpseudocode}
\usepackage{algorithm}
\usepackage{algorithmicx}
\usepackage{algpseudocode}
\usepackage{verbatim}
\usepackage{graphicx}
\usepackage{multirow}
\usepackage{amsfonts}
\usepackage{cite}
\usepackage{multicol}
\usepackage{balance}

\def\BibTeX{{\rm B\kern-.05em{\sc i\kern-.025em b}\kern-.08em
    T\kern-.1667em\lower.7ex\hbox{E}\kern-.125emX}}
\usepackage{balance}
\begin{document}
\title{FedEx: Expediting Federated Learning over Heterogeneous Mobile Devices by Overlapping and Participant Selection}
\author{Jiaxiang Geng,~\IEEEmembership{Student Member,~IEEE}, Boyu Li, Xiaoqi Qin,~\IEEEmembership{Senior Member,~IEEE}, Yixuan Li,~\IEEEmembership{Student Member,~IEEE}, Liang Li,~\IEEEmembership{Member,~IEEE}, Yanzhao Hou,~\IEEEmembership{Member,~IEEE}, and Miao Pan,~\IEEEmembership{Senior Member,~IEEE} \vspace{-11mm}
\thanks{J. Geng, B. Li, X. Qin and Y. Hou are with the State Key Laboratory of Networking and Switching Technology,  Beijing University of Posts and Telecommunications, Beijing, 100876, China, (e-mail: \{lelegjx, liboyu, xiaoqiqin, houyanzhao\}@bupt.edu.cn).

Y. Li is with the College of Electronic Information Engineering, Taiyuan University of Technology, Taiyuan, 030024, China (e-mail: liyixuan@tyut.edu.cn).

L. Li is with the Department of Advanced Interdisciplinary Research, Pengcheng Laboratory, 518055, China, (e-mail: lil03@pcl.ac.cn).

M. Pan is with the Department of Electrical and Computer Engineering, University of Houston, Houston, TX 77204 USA, (e-mail: mpan2@uh.edu).

This work of Y. Hou was supported in part by the Joint Research Fund for Beijing Natural Science Foundation and Haidian Original Innovation under Grant L232001, in part by the National Natural Science Foundation of China under Grant 62327801, in part by the Fundamental Research Funds for the Central Universities under Grant 2242022k60006 and in part by the 111 Project of China (No. B16006). This work of X. Qin was supported in part by the National Key Research and Development Program of China under Grant 2023YFB4301901, NSFC Project 62371072 and U22B2003. This work of L. Li was supported in part by the National Natural Science Foundation of China under Grant 62201071. This work of M. Pan was supported in part by the U.S. National Science Foundation under Grant CNS-2431596, CSR-2403249, CNS-2318664 and CNS-2107057.

(Corresponding Author: Xiaoqi Qin, Yanzhao Hou and Miao Pan)}}


\maketitle

\begin{abstract}
Training latency is critical for the success of numerous intrigued applications ignited by federated learning (FL) over heterogeneous mobile devices. By revolutionarily overlapping local gradient transmission with continuous local computing, FL can remarkably reduce its training latency over homogeneous clients, yet encounter severe model staleness, model drifts, memory cost and straggler issues in heterogeneous environments. To unleash the full potential of overlapping, we propose, FedEx, a novel \underline{fed}erated learning approach to \underline{ex}pedite FL training over mobile devices under data, computing and wireless heterogeneity. FedEx redefines the overlapping procedure with staleness ceilings to constrain memory consumption and make overlapping compatible with participation selection (PS) designs. Then, FedEx characterizes the PS utility function by considering the latency reduced by overlapping, and provides a holistic PS solution to address the straggler issue. FedEx also introduces a simple but effective metric to trigger overlapping, in order to avoid model drifts. Experimental results show that compared with its peer designs, FedEx demonstrates substantial reductions in FL training latency over heterogeneous mobile devices with limited memory cost.
\end{abstract}


\vspace{-6mm}
\section{Introduction}
\vspace{-1mm}
Nowadays, the landscape of federated learning (FL) has undergone a significant shift, which has transitioned from its traditional realm within data centers to embrace the capabilities of mobile devices \cite{li2020talk}. This transition is supported by the continuous evolution of hardware, propelling mobile devices such as the NVIDIA Xavier, Galaxy Note20, iPad Pro, MacBook laptops, and UAVs into possessing increasingly potent on-device computing capabilities tailored for localized model training~\cite{guo2022joint}. Adhering to the core FL principle of conducting localized training of deep neural networks (DNNs) while safeguarding the privacy of raw data, the application of FL to mobile devices has ushered in a host of promising possibilities spanning diverse domains including computer vision~\cite{marques2020machine}, natural language processing~\cite{hartmann2019federated} and many other applications such as e-Healthcare~\cite{brisimi2018federated}, voice assistants in smart home~\cite{Nguyen2021Federated}, and autonomous driving~\cite{ye2020federated}. However, in contrast to GPU clusters with Ethernet connections boasting speeds of 50-100 Gbps~\cite{wen2017terngrad}, wireless transmission bandwidth is rather limited, resulting in potentially significant delays in gradient transmission. Besides, the varied computational capabilities and channel conditions among mobile devices often result in ``straggler'' issues, while the non-independent and identically distributed (non-i.i.d.) local data increases the complexity of training to converge. All these factors result in the high latency of FL training over mobile devices, imposing huge burdens on mobile devices and compromising the performance of corresponding applications.


Different from tons of communication/latency efficient FL papers, the concept of overlapping communication and computing, exemplified by delayed gradient averaging (DGA) in~\cite{zhu2021delayed}, tears down the traditional interleaved workflow of local computing and gradient uploading in FL training and magically ``disappears'' the communication time/phase. The innovative aspect lies in that it enables mobile devices to continue training their local models while transmitting their model updates to FL server, i.e., burying/overlapping the communication phase into continuous local computing one. Since both computing and communication contribute to FL convergence, such a ``continuous computing while transmitting'' overlapping method helps to reduce the total latency of FL training and may fundamentally eliminate FL's communication efficiency issue.

Although overlapping based FL has the potential to accelerate FL training, it introduces model staleness problem and may not work well or work at all in heterogeneous environments (i.e., data/computing/wireless heterogeneity). Briefly, since overlapping lets FL clients use the old global model for ongoing local training, the trained local model parameters become stale. To mitigate the model staleness issue, the update correction mechanism can be integrated in overlapping based FL protocols by using stale averaged gradients for homogeneous FL client devices at the huge cost of their memory storage~\cite{zhu2021delayed}. Heterogeneity among clients' mobile devices makes the memory cost issue more severe and almost diminishes the benefits of overlapping computing and communication. For example, following typical overlapping based FL protocols, fast devices (computing + communications) have to wait for the slowest device (i.e., straggler) to finish an FL training round, and those fast devices, in particular fast computing devices, will undergo many rounds of local training in advance before receiving the new/next round global model for update correction. That will accumulate rounds of gradients to correct, causing significant model staleness, and impose a huge burden on, or even overflow fast mobile devices' memory. The advanced local training of those fast devices can also drive local models towards their respective local optima, potentially resulting in divergence from the objective of the global model (a.k.a \textit{model drift}~\cite{karimireddy2020scaffold}) and hindering FL convergence.

\begin{figure} \centering
 \subfigure[Wireless transmission collision of local model updates. \label{fig:DGA_collision}]
  {\includegraphics[width=0.96\linewidth]{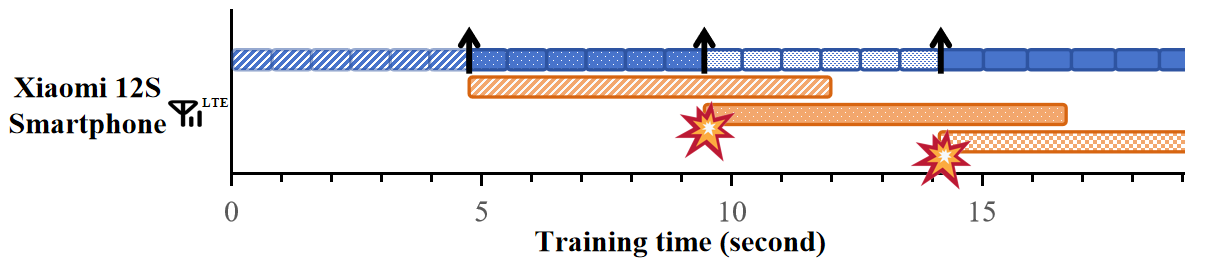}}
 \subfigure[Synchronous (homo.) vs Asynchronous (heter.) local model updates.
 \label{fig:DGA_mismatch}]
{\includegraphics[width=0.96\linewidth]{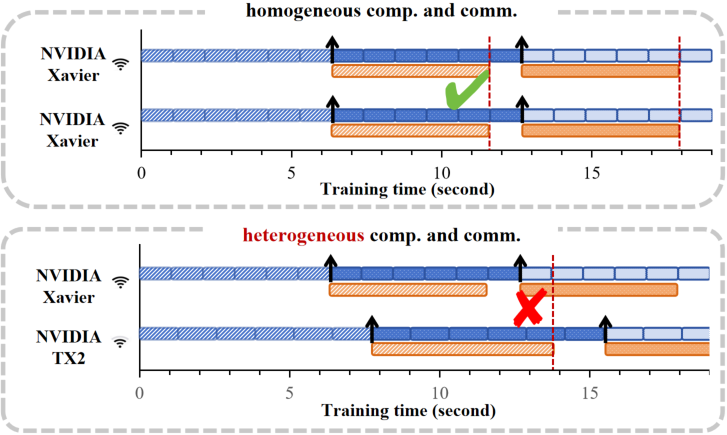}}
{\includegraphics[width=0.96\linewidth]{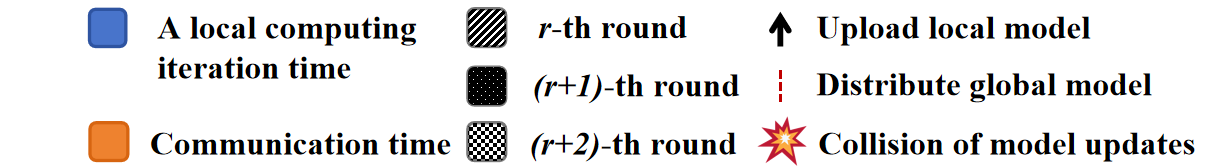}}
\vspace{-2mm}
 \caption{Overlapping deficiencies due to heterogeneity.} \label{fig:DGAHomovsHeter}
\vspace{-6mm}
\end{figure}

Besides, overlapping based FL requires the devices to transmit their local updates when they complete every fixed number of local training iterations, say $K$ iterations. That may result in the ``collision of local updates communications'' for a fast computing and low transmission mobile device (e.g., Xiaomi 12S smartphone with LTE access) in heterogeneous environments, which will fail overlapping based FL. As shown in Fig.~\ref{fig:DGA_collision}, if a device computes $K$ iterations of local training fast and transmit local model updates extremely slow, there will be a collision of communications for local model updates across different rounds, which makes overlapping based FL protocols not work. Moreover, in heterogeneous environments, even though continuous computing may help to speedup FL convergence, the per-round latency is still bottle-necked by the straggler. Thus, there is plenty of room to further improve the delay efficiency in overlapping based FL training.

To alleviate the straggler issue in heterogeneous environments, researchers have explored to select a subset of devices to participate in FL training over rounds, i.e., FL participation selection (PS)~\cite{wang2021device,nishio2019client,fraboni2021clustered,luo2022tackling}. For example, Oort~\cite{lai2021oort}, a pioneering FL PS design, jointly considers the statistical utility (i.e., FL learning convergence/accuracy) and global system utility (i.e., training latency per round). Oort selects participating mobile devices that have good contributions to FL learning convergence and are capable of completing local training before a setup latency threshold. However, existing FL PS methods still follow traditional computing and communication interleaving FL training protocol and don't consider the training speedup brought by continuous computing or overlapping.

Intuitively, binding FL PS designs (e.g., Oort) with overlapping based FL framework may remarkably accelerate FL training, since it (i) mitigates straggler issue and reduces the training latency per round and (ii) takes the latency reduction by overlapping computing and communications into account. But such a simple combination confronts multifaceted challenges. Firstly, it is challenging to identify the ending/starting time point of an FL training round for PS in heterogeneous environments, since heterogeneity among mobile devices transits the overlapping based FL protocol from synchronous local model updates into asynchronous one. As shown in Fig.~\ref{fig:DGA_mismatch}, following the principle of communicating local model updates after $K$ iterations of local computing, fast devices will not wait for the straggler to wrap up the current round, but go ahead to transmit the model updates for the next round while the straggler is still struggling to complete its current round of training. Such asynchronous model updates of overlapping based FL in heterogeneous environments not only lead to model staleness and drift, but also make PS not work and FL convergence intractable. Secondly, it makes no sense to directly use conventional PS utility functions for overlapping based FL, since it doesn't consider the latency reduction brought by the communication and computing overlapping. Thirdly, it tends to prolong convergence time and cause a waste of mobile devices' resources to initiate communication and computing overlapping right after the beginning of FL training. At the early stages of training, the local models trained on heterogeneous data can display significant divergence, and thus the overlapping and continuous training may not help the training converge to a uniform global model. This issue will be further amplified when applying PS in an overlapping based FL framework, where only a subset of devices participates in global aggregation in every round rather than collecting model updates from all devices. Due to data heterogeneity, the selected model updates may be biased, which can slow down convergence and waste mobile devices' computing resources for unnecessary continuous training and overlapping.
To address the challenges above, in this paper, we propose a novel \underline{fed}erated learning approach over heterogeneous mobile devices via overlapping and participant selection to \underline{ex}pedite FL training (FedEx), launching the foundation for overlapping based FL training over heterogeneous devices. FedEx redefines the overlapping procedure, characterizes the PS utility function by considering overlapping benefits, and provides a holistic solution to effectively integrate PS into overlapping based FL framework under data, computing and wireless heterogeneity. It also frees fast mobile devices from being stuck in endless yet unproductive continuous local computing that has no contributions to the FL convergence, thus further improving delay efficiency and resource utilization for FL training over mobile devices. Our salient contributions are summarized as follows.

\begin{itemize}
    \item We customize the overlapping protocol with staleness ceiling for FL training in heterogeneous environments, i.e., overcoming ``collision of local model updates'' and ``asynchronous updates'' issues, and make it compatible with FL PS designs. 
    \item We devise a novel PS utility function to alleviate the straggler effects induced by heterogeneity, thus notably reducing the per round latency, while promoting the benefits of overlapping and continuous computing. 
    \item We propose a local-global model closeness based threshold method to determine when FL training should transit from classical computing and communication interleaved manner into overlapping computing and communication one, which helps to avoid model drifts at early FL training stages. 
    \item We develop a FedEx prototype and evaluate its performance with extensive experiments. The experimental results demonstrate that FedEx can remarkably reduce the latency for FL training over heterogeneous mobile devices at limited costs. Our code has been open-sourced at: https://github.com/BESTTOOLBOX/FedEx.
\end{itemize}

\begin{table}\centering
   \caption{Comp. vs Comm. (Squeezenet@CIFAR10).
   \label{table:compvscomm}}
\small
\vspace{-3mm}
\begin{tabular}{c|c|c|c}
\hline
Mobile Device Configuration\footnotemark[1] & $T_n^{cp}$ & $T_n^{cm}$ & $T_n^{cm}/T_n^{cp}$ \\ \toprule
Xavier w/ Wi-Fi (6.9Mbps) & 1.13s & 5.54s & 4.90 \\ \midrule
TX2 w/ Wi-Fi (6.0Mbps) & 1.35s & 6.40s & 4.74 \\ \midrule
Xiaomi 12S w/ LTE (5.0Mbps) & 0.84s & 7.66s & 9.12 \\ \bottomrule
\end{tabular}
\vspace{-6mm}
\end{table}

\footnotetext[1]{The transmission data rates displayed in Table~\ref{table:compvscomm} represent the averaged achievable data rates for uploading transmissions. Figure~\ref{fig:DGAHomovsHeter} employs the same communication settings.}

\vspace{-4mm}
\section{Background and Motivations}\label{sec:motivation}
\subsection{FL on Mobile Devices}
We explore a wireless federated learning system comprising an FL aggregator, which can be an edge server like a base station or gNodeB, alongside a group of $N$ mobile devices acting as FL clients. Each mobile device exhibits heterogeneity in terms of computational capacity, wireless transmission rates, and data distributions. The typical overarching goal of federated learning can be expressed as:
\vspace{-2mm}
\begin{equation}
\small
 \begin{aligned}
 \min_{w\in \mathbb{R}  ^d} F \left ( w \right ):=\sum_{n=1}^N \pi_n F_n \left ( w \right ) \; ,
 \end{aligned}
 \vspace{-2mm}
\end{equation}
where $\pi_n = B_n / \sum_{n=1}^N B_n$ is the aggregation weight of device $n$ and $B_n$ represents the number of samples in the local dataset. Here, $w$ denotes the model with dimensionality $d$, and $F_n(w)$ represents the local loss function of device $n$. In FL round $r$, the classical FL training process unfolds as follows: the server chooses $P$ devices as $\mathcal{P}_r$ from the total pool of $N$ devices. Subsequently, it broadcasts the latest global model gradient to the chosen devices. Each chosen mobile device $n$ updates its local models by running $K$ local training iterations with global model and their local dataset. The updated local models are then uploaded to the server, typically via communication channels such as 4G, 5G, Wi-Fi 5, etc. The server diligently aggregates these uploaded local models to update the global model. This iterative process continues until FL achieves convergence. 
To achieve fast and accurate model training,
the focus is not only on FL learning performance but also on training delay efficiency. 
In a heterogeneous device environment, each device has varying computation and communication latency, 
and the fast devices need to wait for the straggler (i.e., the slowest device). Therefore, the system latency for each device should consist of three components: the communication latency (i.e., $T_n^{cm}$), the computing latency of local computing iterations for $K$ times (i.e., $K \cdot T_n^{cp}$), and the waiting latency (i.e., $T_{n,r}^{wait}$). 


\vspace{-2mm}
\subsection{Deficiencies of Overlapping Based FL in Heterogeneous Environments}\label{sec:defDGA}
In overlapping based FL,
since the continuous local computing is based on old global model (i.e., the global model from the previous round) instead of the updated one, it leads to the model staleness issue. Here, we denote the ``overlapping staleness'' of device $n$ as $S_n$, which can be expressed as $S_n = \left\lceil (T_n^{cm} + T_n^{wait})/T_n^{cp} \right\rceil$, i.e., the number of iterations for continuous local computing. Current FL training round's overlap staleness is determined by the device with largest staleness level, i.e., $S=\max_n \{S_n\}$. To handle overlapping staleness, some model update correction mechanisms are usually used by replacing the local gradients, which might be several iterations before, with stale averaged gradients. Although overlapping based FL may work well for homogeneous clients, it exhibits many deficiencies and may fail FL over heterogeneous mobile devices. In heterogeneous environments, overlapping based FL may collapse due to the \textit{collision of local model updates}, and even though it can proceed, it is deficient in handling (i) \textit{huge memory cost for update correction}, (ii) \textit{model drift}, and (iii) \textit{straggler issues}.

\begin{figure} \centering
 \subfigure[Homogeneity: only NVIDIA TX2 w/ 30Mbps Wi-Fi. \label{fig:DGA_homogeneous}]
  {\includegraphics[width=0.49\linewidth]{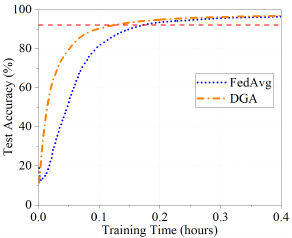}}
 \subfigure[Heterogeneity: NVIDIA TX2 w/ 30Mbps Wi-Fi, Xavier w/ 30Mbps Wi-Fi and Xiaomi 12S w/ 20Mbps LTE. \label{fig:DGA_heterogeneous}]
  {\includegraphics[width=0.49\linewidth]{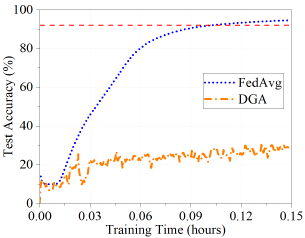}}
\vspace{-5mm}
 \caption{Performance comparison: homo. vs heter. (CNN@MNIST, non-i.i.d. data).} \label{fig:DGAHomovsHeter}
\vspace{-5mm}
\end{figure}

Specifically, following DGA procedure\footnotemark[2]\footnotetext[2]{Please refer to ``Algorithm 1 Delayed Gradient Averaging (DGA)'' in~\cite{zhu2021delayed}.}, the device transmits its local updates when it completes every fixed number of local training iterations, say $K$ iterations. Nevertheless, it may lead to ``collision of local update communications'' issues, if certain device's transmission latency is much larger than its computing time consumption as illustrated in Fig.~\ref{fig:DGA_collision}. Even if the mobile device can be equipped with multiple antennas allowing for simultaneous transmissions of local updates over different rounds, it will transmit the stale local model gradients, i.e., locally trained based on the ($r-1$)-th round global model, for the ($r+1$)-th round updates before it receives the $r$-th round global model for staleness correction. We conducted empirical studies with popular mobile devices and accessing technologies, and the results in Table~\ref{table:compvscomm} show that the fast computing and slow transmission situations largely exist in practice, which can trigger the \emph{collision of communications for local model updates}.
Assuming the overlapping training can proceed without ``collision of local model updates'' in heterogeneous environments, \textit{the memory cost issue for model update correction will be aggravated due to computing and wireless heterogeneity among mobile devices} and \textit{overlapping model staleness may incur model drift issue due to data heterogeneity}, which can degrade or even fail FL convergence. Overlapping based FL requires each device to locally store multiple copies of its old local model updates for update correction, and the memory consumption of the mobile device increases linearly with the staleness level. Different from homogeneous environments, computing and wireless heterogeneity enlarges the time consumption gap between fast devices and the slowest device (i.e., straggler) to complete their local computing and model updates. The enlarged gap directly increases the staleness level of fast devices. When the overlap staleness $S_n$ is large, the heavy gradient storage requirement will impose a huge burden on, or even overflow fast mobile devices' limited memory. Moreover, due to data heterogeneity, the local trained models point to the direction towards the local optima, which may not be consistent with the objective of the aggregated global model, called \textit{model drift}~\cite{karimireddy2020scaffold}. Thus, a high overlapping staleness level, i.e., performing too much continuous local computing, is detrimental to the learning accuracy and may significantly slow down or fail FL convergence under data heterogeneous scenarios. Fig.~\ref{fig:DGAHomovsHeter} shows our empirical studies to compare the testing accuracy of DGA and FedAvg in homogeneous and heterogeneous environments, respectively, which supports our model drift and staleness analysis above.
Besides, although existing overlapping based FL designs benefit from continuous training and accelerate FL training, \emph{the per-round latency can be largely bottle-necked by the straggler in heterogeneous environments}. To achieve good learning performance and low latency, a heterogeneity-aware overlapping design is in need.

\vspace{-3mm}
\subsection{Limitations of SOTA PS Methods}\label{sec:OortLimit}
\vspace{-1mm}
To mitigate the straggler issue in heterogeneous environments, various FL PS designs have been proposed to select a subset of devices to participate in FL training over rounds~\cite{wang2021device,nishio2019client,fraboni2021clustered,luo2022tackling}. As an advanced FL PS design, Oort~\cite{lai2021oort} novelly considers both the statistical utility (i.e., FL learning convergence/accuracy) and global system utility (i.e., training latency per round). The utility function of device $n$ in Oort can be expressed as follows.
\vspace{-1mm}
\begin{equation}
\small
\begin{split}
Util(n) = |B_n|\sqrt{\frac{1}{|B_n|}\sum_{\substack{k\in {B}_n}}Loss(k)^2} 
\times (\frac{T}{t(n)})^{\mathbb{I}(T<t(n)) \times \alpha}  \; ,
\end{split}
\label{oortUtil}
\vspace{-2mm}
\end{equation}
where $B_n$ represents the local training samples of mobile device $n$, $Loss(k)$ corresponds to the training loss of the $k$-th sample, $t(n)$ signifies the system latency of mobile device $n$, $T$ is the developer-preferred duration for each round, and $\alpha$ serves as the penalty factor. Additionally, $\mathbb{I}(x)$ denotes an indicator function, where $\mathbb{I}(x)=1$ when $x$ is true, and $0$ otherwise. Under Eqn.~(\ref{oortUtil}), top $K$ devices with the highest utility will be selected for FL training in synchronous manner. Instead of picking devices with top-$K$ utility deterministically over rounds, Oort employs a ``temporal uncertainty'' mechanism~\cite{lai2021oort} to gradually increase the utility of a device if it has been overlooked for a long time. Thus, low-utility devices still have chances to be selected to avoid biased FL training.

Despite the impressive performance, existing PS designs (e.g., Oort) still follow traditional computing and communication interleaving FL training protocol, which results in devices having to wait for uploading of local updates (i.e., $T_n^{cm}$) and straggler (i.e., $T_{n,r}^{wait}$), once they have finished local computing. In particular, for FL training in computing and wireless heterogeneous environments, the waiting time for fast computing devices can be very long. \textit{If the wasted waiting time can be effectively utilized, the latency of PS based FL training has the potential to be further reduced.} That gives the opportunities to investigate the binding of overlapping and PS to push the boundaries of FL performance.

\vspace{-3mm}
\subsection{Challenges in Overlapping + PS
}\label{sec:HammeringDGAOort}
\vspace{-1mm}
Although directly binding overlapping and PS is promising to accelerate FL training, this straightforward combination faces multifaceted challenges.

Firstly, \textit{overlapping induced asynchronous local model updates conflict with synchronous PS in heterogeneous environments.} Again, each device will transmit its local model updates after $K$ iterations of local computing in DGA. Considering computing and wireless heterogeneity among mobile devices, synchronous local model updates cannot be guaranteed anymore, since fast devices may already initiate the communications of the next-round model updates while the straggler is still struggling to wrap up the current round training as shown in Fig.~\ref{fig:DGA_mismatch}.
However, pioneering PS designs, such as Oort, are commonly tailored for synchronous FL training, where FL server uniformly evaluates the utility of all candidate devices at the beginning of each round for PS. 
The asynchronous model updates in overlapping based FL framework essentially conflict with the synchronous FL training requirements of PS, which makes directly combining them impossible.

Secondly, \emph{PS utility function doesn't take the latency reduction brought by overlapping into account}. Simply, the overlapping and continuous training changes the way of traditional computing and communication interleaved FL training, which consequently affects the definition of computing latency in a training round and corresponding per round training latency for a candidate mobile device. While the \emph{global system utility}~\cite{lai2021oort} in Oort's utility function is based on conventional per round FL training latency, directly using Oort to select the participating mobile devices for overlapping based FL is not appropriate, since it is not aware of overlapping benefits to accelerate FL training convergence.

Thirdly, \emph{it may cause model drift and waste of resources for mobile devices to activate overlapping training while conducting PS at very early FL training stages}. Starting from scratch, local models trained on heterogeneous data may exhibit significant differences. At very early FL training stages, overlapping and continuous computing can potentially exacerbate the training towards local optima, which may hinder the training converges to a unified global model. 
Moreover, due to data heterogeneity, selecting only a subset of mobile devices to participate in overlapping training may intensify biased training and further slow down FL convergence rate. Besides, the unnecessary overlapping and continuous training above wastes the valuable computation and memory resources of mobile devices. Therefore, it is worthy to study when to trigger the overlapping, in order to facilitate FL training over heterogeneous mobile devices.

\vspace{-2mm}
\section{FedEx Design}
\vspace{-1.5mm}
\subsection{FedEx Overview}
Motivated by the deficiencies/limitations of overlapping/PS and challenges in hammering overlapping and PS, we propose a novel \underline{fed}erated learning approach over heterogeneous mobile devices via overlapping and participant selection to \underline{ex}pedite FL training (FedEx). Our design consists of four coherent parts: (i) model updates with staleness ceiling, (ii) overlapping aware PS utility function, (iii) FedEx overlapping trigger, and (iv) discussion on staleness ceiling. The sketch of FedEx is shown in Fig.~\ref{fig:FedExArch}.

\begin{figure*}
\centering
\includegraphics[width=1\textwidth]{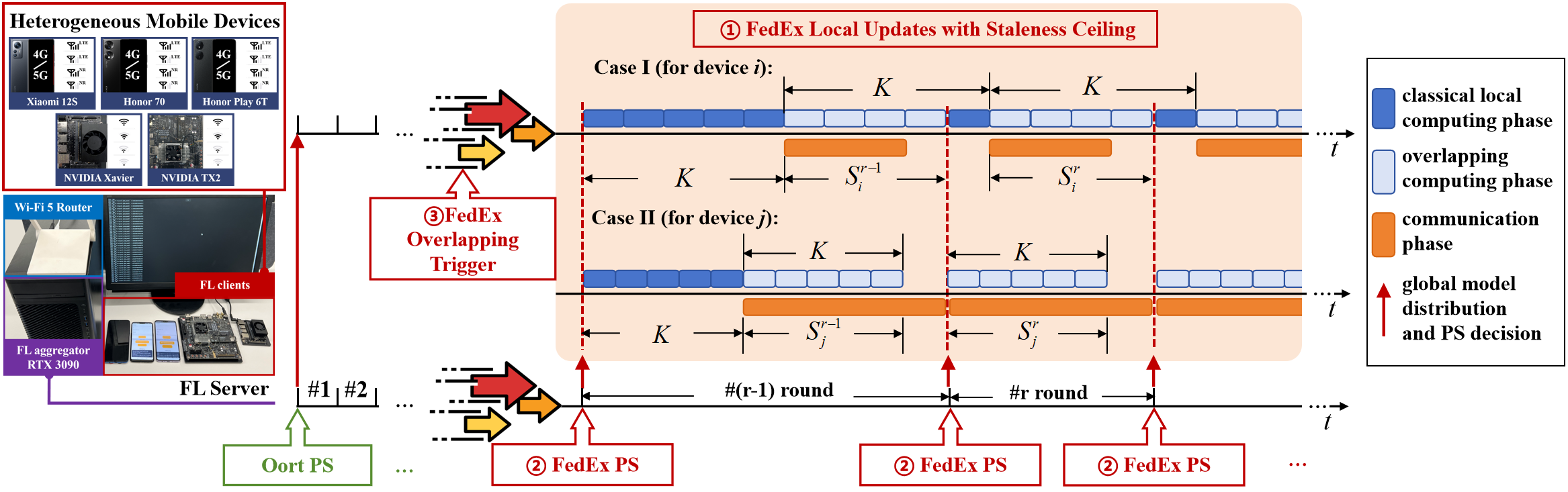}
\vspace{-6mm}
\caption{The sketch of FedEx procedure and testbed.}
\label{fig:FedExArch}
\vspace{-6mm}
\end{figure*}

\vspace{-3mm}
\subsection{Model Updates with Staleness Ceiling}\label{sec:Ceiling}
To preclude the collision of communications, alleviate model drift, constrain the memory cost for continuous computing (Sec.~\ref{sec:defDGA}) and synchronize devices' local model updates in every round (Sec.~\ref{sec:HammeringDGAOort}), we propose a novel local model updates protocol with staleness ceiling in FedEx. As discussed in Sec.~\ref{sec:motivation}, we found that in heterogeneous environments, existing overlapping principle (e.g. DGA) of uploading local gradients after every $K$ computing iterations is the root for the collision of communications and asynchronous model updates, and its endless/non-constrained continuous local computing aggravates model staleness, which results in model drift as well as huge memory consumption for updates correction. So, to address those issues and fit FL training over heterogeneous mobile devices, FedEx redefines the rule of overlapping local model updates and casts a ceiling on staleness/the iteration number of continuous local computing.

For ease of description, let us divide the local computing phase for device $n$ in any FL round $r$ into two sub-phases in FedEx, i.e., classical local computing phase (i.e., $K$) and continuous/overlapping computing one (i.e., $S_n^r$). In particular, FedEx synchronizes the end of current round/start of the next round among heterogeneous devices by the end of straggler's current round training (i.e., the time point of accomplishing classical local computing + model updates transmission), while allowing participating devices for overlapping/continuous computing. Within each FedEx round, the mobile device can only submit its local model updates once after completing its classical local computing phase. At the same time, the device is allowed to conduct its continuous computing before the end of this FL training round as shown in Fig.~\ref{fig:FedExArch}. Compared with existing overlapping principle, although FedEx's rule may delay the local model updates of fast computing devices for the next or next few rounds, it avoids the collision of communications and lays a synchronous FL training foundation for overlapping and PS integration. 

Coupling with the updated model updates protocol, we cast a ceiling on the staleness/the iteration number of continuous computing (i.e., $S_n$). For device $n$ in the $r$-th round, we let $S_n^r = \min \{\lceil (T_n^{cm} + T_{n,r}^{wait})/T_n^{cp} \rceil , K\}$.

For participating devices operating under the staleness ceiling, they may adhere to two possible cases, as depicted in Fig.~\ref{fig:FedExArch}. (1) \textbf{Case I} ($S_n^r < K$): Device $n$, if it is selected to participate, conducts ($K-S_n^{r-1}$) classical local computing iterations in the $r$-th round, and transmits the local updates while proceeding with continuous computing for $S_n^r$ iterations until the end of the $r$-th round (i.e., the end of straggler's training in the $r$-th round). The updated local model after $S_n^r$ iterations continuous computing is stored in memory. (2) \textbf{Case II} ($S_n^r = K$): Device $n$, if it is selected to participate, conducts ($K-S_n^{r-1}$) classical local computing iterations in the $r$-th round, and transmits the local updates while proceeding with continuous computing for $S_n^r = K$ iterations.
After that, it halts its continuous local computations, stores the updated local model in memory, and waits idly until the end of the $r$-th round. 
Note that even though staleness ceiling introduces idle waiting time for fast computing devices, it does help to alleviate model drift and avoid excessive memory usage in the original overlapping principle.

\vspace{-4mm}
\subsection{Overlapping Aware PS Utility Function}\label{sec:FLOODutility}
To address straggler issues for latency reduction per round (Sec.~\ref{sec:defDGA}) and make PS aware of overlapping benefits (Sec.~\ref{sec:OortLimit} and Sec.~\ref{sec:HammeringDGAOort}), we develop an overlapping aware utility function for FL PS in heterogeneous environments.
Inspired by Oort’s utility function~\cite{lai2021oort} in Eqn.~(\ref{oortUtil}), the proposed PS utility function aims to jointly improve statistical utility and delay efficiency of FL training, while considering the latency reduction brought by overlapping computing and communication. Thus, we revise global system utility in Eqn.~(\ref{oortUtil}) into overlapping latency utility in FedEx, and present the overlapping aware PS utility function as follows.
\vspace{-4mm}
\begin{equation}
\vspace{-3mm}
\small
\begin{split}
U(n,r) \!=\! 
\underbrace{|B_n^r|\sqrt{\frac{1}{|B_n^r|}\sum_{\substack{k\in {B}_n^r}}\!Loss(k)^2}}_{\textbf{Statistical utility}} \!\times\! \underbrace{\left (\frac{1}{(K \!-\! S_n^{r - 1}) \!\cdot\! t_n^r \!+\! \frac{D_n}{R_n^r}}\right )^{\alpha}}_{\textbf{Overlapping latency utility}},
\end{split}
\label{Util}
\vspace{-3mm}
\end{equation}
where $|B_n^r|$ is the number of training data samples of mobile device $n$ in the $r$-th round, $t_n^r$ is the averaged time consumption for one iteration of local computing at mobile device $n$~\cite{RuiEEFL}, $D_n$ is the model size of client $n$ and $R_n^r$ is the transmission rate of mobile device $n$ in the $r$-th round. $\frac{D_n}{R_n^r}$ represents the mobile device $n$'s estimated transmission time. Here, $\alpha$ is the scaling coefficient to trade off the statistical utility and overlapping latency utility.

The utility function above consists of two parts: statistical utility and overlapping latency utility. The statistical utility accounts for the estimated learning performance contributions to FL convergence, which is the same as that in Oort~\cite{lai2021oort}.  
The overlapping latency utility has two components: classical computing latency and the model update transmission latency. For classical computing latency in the $r$-th round, mobile device $n$ needs to complete $K$ local training iterations. Regarding the latency reduction by overlapping, device $n$ has already conducted continuous computing for $S_n^{r-1}$ iterations in the ($r-1$)-th round. Thus, in the $r$-th round, device $n$ only needs to perform ($K - S_n^{r-1}$) iterations for classical computing phase before transmitting its local gradients. As illustrated in~\cite{RuiEEFL}, given a learning task and the computing capability of mobile device $n$, $t_n^r$ is relatively fixed in practice and can be estimated with good precision by averaging the latency for the previous iterations. Therefore, device $n$'s classical computing latency in the $r$-th round can be estimated as $(K - S_n^{r - 1}) \cdot t_n^r$. For transmission latency, given the fact that mobile devices' wireless communication conditions don't significantly change within a short period of time, we assume $R_n^r$ is fixed in the $r$-th round and can be approximated by using $n$'s transmission rate at the beginning of this round.

The proposed overlapping latency utility is aware of overlapping benefits by considering latency reduction $(K - S_n^{r - 1}) \cdot t_n^r$. Besides, as discussed in Sec.~\ref{sec:Ceiling}, the per-round latency in FedEx is determined by the straggler's training latency, i.e., the slowest device's classical on-device training time plus its communication time for transmitting local updates. The overlapping latency utility helps to address straggler issues since it will select fast devices with small training latency in the round, i.e., $[(K - S_n^{r - 1}) \cdot t_n^r + \frac{D_n}{R_n^r}]$. Together with statistical utility, the proposed overlapping aware utility function\footnotemark[3]\footnotetext[3]{Regarding the privacy preservation of statistical utility and overlap latency utility, we leverage the same privacy analysis as presented in Oort \cite{lai2021oort}.} will select the mobile devices with good learning contributions and high overlapping latency utility to participate, which can accelerate the FL training over heterogeneous mobile devices\footnotemark[4]\footnotetext[4]{For device $n$ not selected to participate in the $r$-th round, it refrains from any computing and transmission activities. Instead, it inherits the values of $K$ and $S_n^{r-1}$ from the $(r-1)$-th round, updates them into $K$ and $S_n^{r}$, and wait for the next time to participate. For devices overlooked for a long time, we follow the ``temporal uncertainty'' mechanism to gradually increase their utilities, in order to avoid biased training in heterogeneous environments. Due to page limits, the duplicated discussions have been omitted here.}. 

\vspace{-2mm}
\subsection{FedEx Overlapping Trigger}\label{sec:floodtrigger}
To avoid model drift and save the valuable resources (e.g., computing, memory and battery) of mobile devices at early training stages (Sec.~\ref{sec:HammeringDGAOort}), we propose a simple yet effective metric to trigger overlapping in FedEx.
As discussed in Sec.~\ref{sec:HammeringDGAOort}, local models trained on heterogeneous data may exhibit significant differences at early FL training stages, leading to model drift, and too early overlapping/continuous computing makes it even worse, which wastes mobile devices' resources in terms of computation, memory, and battery. Thus, it is crucial to determine when to trigger the overlapping. Here, we propose to evaluate the level of similarity between local models and the global model, and activate the overlapping when the averaged similarity scores pass a predetermined accuracy threshold. The overlapping trigger is defined as follows.

\vspace{-3mm}
\begin{equation}
\small
\frac{1}{N} \sum_{n=1}^N CKA(w_n^{r-1},w_g^{r-1})>\delta_{CKA},
\label{threshold}
\vspace{-2mm}
\end{equation}
where $\delta_{CKA}$ is the threshold and $0 < \delta_{CKA} < 1$, which varies with different learning tasks, and centered kernel alignment (CKA) is used to measure the similarity of the output features between local models and the global model, which outputs a score between 0 (not similar at all) and 1 (identical). Equation~(\ref{threshold}) indicates that when the averaged CKA value of $N$ participating devices exceeds the threshold, signifying a certain level of similarity between local models and the global model is achieved, the overlapping in FedEx will be triggered to execute. Prior to that, the FL training at early stages follows Oort's approach, avoiding model drift, saving the resources of mobile devices, and preparing for the overlapping in FedEx as shown in Fig.~\ref{fig:FedExArch}.
\vspace{-4mm}
\subsection{Generalization of Staleness Ceiling}\label{sec:generalization_ceiling}


As illustrated in Sec.~\ref{sec:Ceiling}, FedEx\footnotemark[5]\footnotetext[5]{It is noteworthy that DGA can be regarded as a special case of FedEx with $S_n^r\leq K$, where all participating devices are homogeneous.} allows device $n$ to perform at most $K$ more continuous computing iterations, i.e., $S_n^r \leq K$ in the current FL round $r$, and store a copy of the computed local model to correct for the next round. To better understand the impact of FedEx's staleness ceiling, in this subsection, we generalize the overlapping staleness ceiling from $S_n^r \leq K$ to $S_n^r = a_n^r*K+b_n^r$, where $a_n^r$ and $b_n^r$ are two non-negative integer parameters. Based on the generalized staleness ceiling, device $n$ has to store $a_n^r$ copies of the continuous-computed local models, correct them and upload them sequentially in the next $a_n^r$ FL training rounds. Denote the upper bound of \( S_n^r \) by \( U \), where \(U \leq \lceil (T_n^{cm} + T_{n,r}^{wait})/T_n^{cp} \rceil \).

Note that $S_n^r$ plays a crucial role in balancing the pros (i.e., FL training speedup) and cons (i.e., model drift, which slows down FL convergence) of overlapping/continuous computing. Specifically, if $S_n^r$ is smaller, the continuous computing phase is shorter, resulting in fewer local gradient descents within a certain period. This leads to a smaller acceleration effect from overlapping but also mitigates model drift. Conversely, if $S_n^r$ is larger, the continuous computing leads to larger acceleration benefits while incurring bigger model drift. Therefore, $S_n^r$ should tradeoff model drift and overlapping acceleration in order to reduce the FL training latency.


From the perspective of memory consumption, as $S_n^r$ increases, $a_n^r$ monotonically increases, so that mobile device $n$ has to store more copies of staled local models with additional memory cost. That may result in excessive memory consumption, or even memory overflow, especially when device $n$ is much faster than the straggler device. By contrast, when $S_n^r$ is set to be less than certain threshold, say $K$, it ensures that only limited copies (one for $S_n^r = K$) of the local models are stored, which potentially avoids excessive memory consumption issue. Experimental results will also be presented in Sec.~\ref{sec:ExpCeiling} to corroborate our analysis.

\vspace{-3mm}
\section{Experimental Setup}
\vspace{-1mm}
\subsection{FedEx Implementation}

The FedEx system has been implemented on a testbed, featuring an FL aggregator and a set of real-world heterogeneous mobile devices, detailed in the Appendix E. FedEx is implemented by building on top of FLOWER~\cite{beutel2021flower}. 
In order to realize FedEx, 
we need to obtain relevant parameters, such as transmission rate and training latency. 
To estimate on-device transmission rates, 
we employed the Network Monitor toolbox in the Android kernel on the smartphone, and self-developed rate monitoring module for NVIDIA devices. 
To measure the latency, the time of model training and transmission are recorded on mobile devices.

\begin{figure*} \centering
 \subfigure[CNN@MNIST. \label{fig:cnn}]
  {\includegraphics[width=0.24\linewidth]{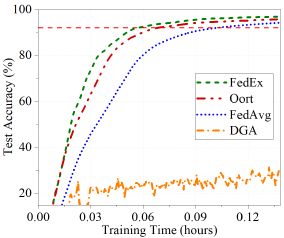}}
 \subfigure[LSTM@Shakespeare. \label{fig:lstm}]
  {\includegraphics[width=0.24\linewidth]{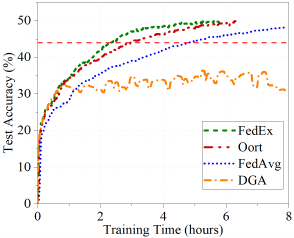}}
  \subfigure[SqueezeNet@CIFAR10. \label{fig:squeezenet}]
  {\includegraphics[width=0.24\linewidth]{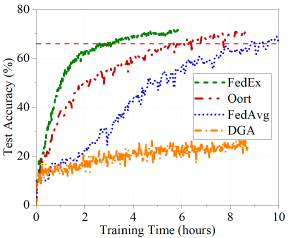}}
  \subfigure[CNN@HAR. \label{fig:har}]
  {\includegraphics[width=0.24\linewidth]{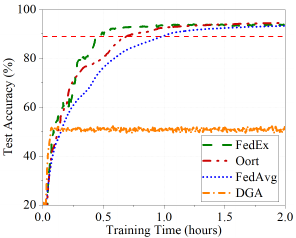}}
\vspace{-2mm}
 \caption{Performance comparison in terms of testing accuracy and training latency under different learning tasks.} \label{fig:overall}
\vspace{-4mm}
\end{figure*}

\vspace{-4mm}
\subsection{Models, Datasets and Parameters}
We evaluate FedEx's performance for three different FL tasks: image classification, next word prediction, and human activity recognition. 
We use four DNN models: a 2-layer CNN~\cite{pmlr-v54-mcmahan17a}, SqueezeNet~\cite{2016arXiv160207360I}, MobileFormer~\cite{9880374}, and LSTM~\cite{hochreiter1997long}, 
where MobileFormer is an efficient model that bridges MobileNet and Transformer. 

As for the image classification task, 
we train a 2-layer CNN on MNIST dataset~\cite{deng2012mnist}, a SqueezeNet on CIFAR10 dataset~\cite{krizhevsky2009learning} and a MobileFormer on TSRD dataset~\cite{TSRD}. 
The MNIST dataset comprises 10 categories, ranging from digit ``0'' to ``9'', and encompasses a total of 60,000 training images and 10,000 validation images.
The CIFAR10 dataset comprises 50,000 training images and 10,000 validation images, distributed across 10 categories. 
The TSRD dataset contains 6164 traffic sign images, representing 58 sign categories. 
As for data distribution among mobile devices, 
we denote $\lambda$ as the non-i.i.d. levels, where $\lambda=0$ indicates that the data among mobile devices is i.i.d., 
$\lambda=0.5$ indicates that 50\% of the data belong to one label and the remaining belong to other labels, 
and $\lambda=1$ indicates that each mobile device owns a disjoint subset of data with one label.
As for the next word prediction task, 
we train an LSTM on the Shakespeare dataset~\cite{Shakespeare}, which consists of 1,129 roles and each of which is viewed as a device.
Since the number of lines and speaking habits of each role vary greatly, the dataset is non-i.i.d. and unbalanced.
As for the human activity recognition task, 
we train a 2-layer CNN on the HAR dataset~\cite{Stisen2015SmartDA}, which is generated by having volunteers wear Samsung Galaxy S2 smartphones equipped with accelerometer and gyroscope sensors. 
The HAR dataset includes 10,299 data samples consisting of six categories of activities, i.e., walking, walking upstairs, walking downstairs, sitting, standing, and laying. 
The dataset is non-i.i.d. due to the different behavioral habits of the volunteers.

We use the following default parameters unless specified otherwise. 
We set the data distribution parameter $\lambda=0.5$. 
The scaling coefficient in the proposed FedEx utility function is set as $\alpha=2$. 
The FedEx threshold is set as $\delta_{CKA}=0.7$. The number of local computing iterations is set as $K=10$.  The upper limit of the staleness ceiling is set as $U=K=10$.

\vspace{-4mm}
\subsection{Baselines for Comparison}
We compare FedEx with the following peer FL designs under different FL tasks.\\
\textbf{FedAvg~\cite{mcmahan2017communication}}: The FL server selects participating devices randomly, and the selected devices perform a fixed number of local training iterations.  \\
\textbf{Oort~\cite{lai2021oort}}: The FL server selects participating devices based on the Oort utility function as defined in Eqn.~(\ref{oortUtil}), and the selected devices perform a fixed number of local iterations.\\
\textbf{DGA~\cite{zhu2021delayed}}: The FL server selects all mobile devices to participate in each round. The selected devices employ an overlapping training mechanism where local training continues while transmitting the model updates (the transmission starts after a fixed number of local training iterations\footnotemark[6]\footnotetext[6]{Note that Oort/DGA is the well-recognized and top design, if not the best, in FL participant selection/overlapping FL domain, respectively. Either of them outperforms several baselines as shown in their own papers~\cite{lai2021oort,zhu2021delayed}. If the proposed FedEx performs better than DGA/Oort, it indicates that FedEx also surpasses those baselines in Oort/DGA paper.}).\\
\textbf{DGAplus}: The FL server selects participating devices randomly. The selected devices employ DGA policy (overlapping training) with staleness ceiling as proposed in Sec.~\ref{sec:Ceiling}. \\
\textbf{DGAplus-Oort}: The FL server selects participating devices based on the Oort utility function as defined in Eqn.~(\ref{oortUtil}). The selected devices employ DGA policy (overlapping training) with staleness ceiling as proposed in Sec.~\ref{sec:Ceiling}.


\vspace{-2mm}
\section{Evaluation Results \& Analysis}
\subsection{Advantages in FL Training Acceleration}

\begin{table*}[]
\centering
   \caption{Overall result of FL training with non-i.i.d. data to reach the target testing accuracy. \\ (OL: \underline{O}verall \underline{L}atency (h), NR: \underline{N}umber of \underline{R}ounds, PRT: Average \underline{P}er \underline{R}ound \underline{T}ime (h), SU: \underline{S}peed\underline{u}p)}
   \label{table:comparisonSOTA}
\small
\setlength{\tabcolsep}{2pt}
\vspace{-2mm}
\resizebox{1\textwidth}{!}{
\begin{tabular}{c|cccccccccccc|cccc|cccc}
\toprule
\multirow{2}{*}{Task} & \multicolumn{12}{c|}{CV} & \multicolumn{4}{c|}{NLP} & \multicolumn{4}{c}{HAR} \\ \cline{2-21} 
 & \multicolumn{4}{c|}{CNN@MNIST} & \multicolumn{4}{c|}{SqueezeNet@CIFAR10} & \multicolumn{4}{c|}{MobileFormer@TSRD} & \multicolumn{4}{c|}{LSTM@Shakespeare} & \multicolumn{4}{c}{CNN@HAR} \\ \hline
Target Acc & \multicolumn{4}{c|}{92\%} & \multicolumn{4}{c|}{66\%} & \multicolumn{4}{c|}{48\%} & \multicolumn{4}{c|}{44\%} & \multicolumn{4}{c}{89\%} \\ \hline
Methods & OL & NR & PRT & \multicolumn{1}{c|}{SU} & OL & NR & PRT & \multicolumn{1}{c|}{SU} & OL & NR & PRT & SU & OL & NR & PRT & SU & OL & NR & PRT & SU \\ \hline
FedAvg & 1.03x$10^{-1}$ & 96 & 1.07x$10^{-3}$ & \multicolumn{1}{c|}{1.0x} & 8.87 & 409 & 2.17x$10^{-2}$ & \multicolumn{1}{c|}{1.0x} & 79.37 & 332 & 2.39x$10^{-1}$ & 1.0x & 4.81 & 174 & 2.77x$10^{-2}$ & 1.0x & 1.05 & 155 & 6.45x$10^{-3}$ & 1.0x \\ \hline
Oort & 6.85x$10^{-2}$ & \textbf{79} & 8.67x$10^{-4}$ & \multicolumn{1}{c|}{1.5x} & 5.98 & \textbf{342} & 1.75x$10^{-2}$ & \multicolumn{1}{c|}{1.5x} & 50.36 & \textbf{215} & 2.34x$10^{-1}$ & 1.6x & 2.96 & 176 & \textbf{1.68x$\mathbf{10^{-2}}$} & 1.6x & 7.58x$10^{-1}$ & \textbf{133} & 5.26x$10^{-3}$ & 1.4x \\ \hline
DGA & \multicolumn{4}{c|}{Max Acc: 31\%} & \multicolumn{4}{c|}{Max Acc: 27\%} & \multicolumn{4}{c|}{Max Acc: 13\%} & \multicolumn{4}{c|}{Max Acc: 36\%} & \multicolumn{4}{c}{Max Acc: 50\%} \\ \hline
\textbf{FedEx} & \textbf{5.74x$\mathbf{10^{-2}}$} & 94 & \textbf{6.11x$\mathbf{10^{-4}}$} & \multicolumn{1}{c|}{\textbf{1.8x}} & \textbf{2.68} & 395 & \textbf{6.78x$\mathbf{10^{-3}}$} & \multicolumn{1}{c|}{\textbf{3.0x}} & \textbf{39.48} & 251 & \textbf{1.57x$\mathbf{10^{-1}}$} & \textbf{2.0x} & \textbf{2.34} & \textbf{118} & 1.98x$10^{-2}$ & \textbf{2.1x} & \textbf{4.87x$\mathbf{10^{-1}}$} & 180 & \textbf{2.78x$\mathbf{10^{-3}}$} & \textbf{2.0x} \\ \bottomrule
\end{tabular}}
\vspace{-6mm}
\end{table*}

As the evaluation results shown in Table~\ref{table:comparisonSOTA} and Fig.~\ref{fig:overall}, the proposed FedEx consistently outperforms its peer designs across various FL tasks, achieving remarkable training speedup. Compared with the baseline, FedAvg, FedEx expedites the FL training to the target test accuracy by approximately 1.8x, 3.0x, 2.0x, 2.1x, and 2.0x for learning tasks including CNN@MNIST, SqueezeNet@CIFAR10, MobileFormer@TSRD, LSTM@Shakespeare, and CNN@HAR, respectively. Notably, FedEx exhibits superior performance to DGA in terms of both testing accuracy and delay efficiency. As discussed in Sec.~\ref{sec:defDGA}, DGA faces significant model staleness, model drift and straggler issues in heterogeneous environments, which impede or prevent the convergence of FL training. Thus, as shown in Table~\ref{table:comparisonSOTA}, DGA only achieves the maximum accuracy rates of 31\%, 27\%, 13\%, 36\% and 50\% for the learning tasks mentioned above. In contrast, FedEx effectively manages overlap staleness and ensures timely global gradient updates. By strategically selecting appropriate devices for participation according to its overlapping aware utility function, FedEx greatly relieves the straggler effects, and achieves much better learning performance with less latency than DGA. While Oort considers both statistical utility (i.e., contributions to FL convergence) and global system utility (i.e., a latency threshold) for PS, the selected fast computing devices have to wait idly until the selected slowest device\footnotemark[7]\footnotetext[7]{It could be one of the fast computing devices, if that mobile device has very low wireless transmission rate.} finishes its current round training (computing + communications) before resuming local training for the next round. 
In contrast, FedEx allows for overlapping and continuous local computing, which improves the utilization of otherwise wasted waiting time, and thus further accelerates FL training.
Consequently, compared with Oort, FedEx achieves a notable speedup of 1.2x, 2.0x, 1.3x, 1.3x, and 1.4x for the tested five learning tasks, respectively.

\begin{figure} \centering
 \subfigure[Mobileformer@TSRD.\label{fig:memory_mobileformer}]
  {\includegraphics[width=0.49\linewidth]{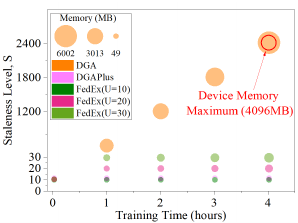}}
  \subfigure[LSTM@Shakespeare.\label{fig:memory_lstm}]
  {\includegraphics[width=0.49\linewidth]{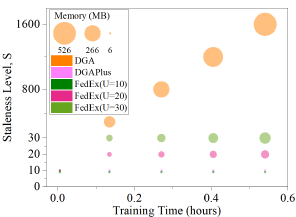}}
\vspace{-5mm}
\caption{Staleness/memory vs FL training time.} \label{fig:memory}
\vspace{-4mm}
\end{figure}

\begin{figure} \centering
 \subfigure[LSTM@Shakespeare.\label{fig:utility_lstm}]
  {\includegraphics[width=0.455\linewidth]{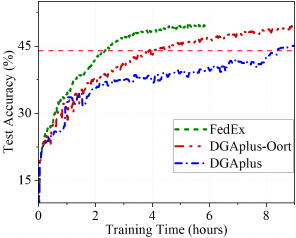}}
  \subfigure[CNN@MNIST.\label{fig:utility_mnist}]
  {\includegraphics[width=0.455\linewidth]{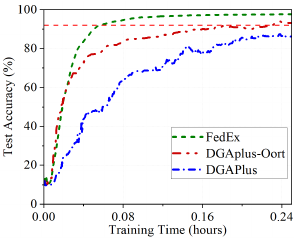}}
  \vspace{-3mm}
  \caption{Performance of overlapping aware PS.} 
  \label{fig:utility}
\vspace{-5mm}
\end{figure}

\vspace{-2mm}
\subsection{Advantages of the Staleness Ceiling}\label{sec:ExpCeiling}
Figure~\ref{fig:memory} illustrates the effectiveness of FedEx with staleness ceiling in reducing model staleness ($S$) and memory usage. Here, the position of the dot reflects the model staleness level at the corresponding training time, and the size of the dot represents the memory usage. In the context of the Mobileformer@TSRD FL task, we compare FedEx with DGA and DGAplus. Figure~\ref{fig:memory} reveals an aggressive increase of overlapping staleness levels in DGA when FL training proceeds. Specifically, staleness rises from $S=10$ at $t=0.03h$ to $S=605$ at $t=1.0h$, and further escalates to $S=2420$ by $t=4.0h$. This rise in overlapping staleness results in substantial memory consumption, averaging at $49.4MB$, $1482.0MB$, and $6002.1MB$ at $t=0.03h$, $t=1.0h$, and $t=4.0h$, respectively. Especially, at $t=4.0h$, where the averaged memory usage surpasses the 4096MB limit of the NVIDIA Jetson TX2, there is a critical risk of memory overflow, which may cause premature termination of FL training. As FL training further proceeds, the model staleness becomes more severe and overflows the memory of more mobile devices, which hinders FL convergence towards the desired accuracy. For instance, DGA can only achieve the maximum accuracy of 13\% for the MobileFormer@TSRD task. Notable improvements can be observed in DGAplus/FedEx, which adds the proposed staleness ceiling. Here, overlapping staleness remains constant at $S=10$ throughout FL training, significantly reducing memory consumption to a stable 49.4MB. We have also tried more staleness ceilings (i.e., $S\leq U$ and $U=10,20$ and $30$, respectively) with different tasks (e.g., LSTM@Shakespeare) and shown the results in Fig.~\ref{fig:memory}. The same analysis applies, i.e., the staleness ceiling can well regulate the memory consumption of continuous computing/overlapping when FL training proceeds in heterogeneous environments.

\begin{table}
\centering
   \caption{Advantages of overlapping aware PS (LSTM@Shakespeare). 
   \label{table:Utility}}
\small
\vspace{-2mm}
\begin{tabular}{ccc}
\toprule
\begin{tabular}[c]{@{}c@{}}Methods\\ (Target Acc: 44\%)\end{tabular} & OL(h) & SpeedUp \\ \hline
DGAplus                                                             & 8.4   & 0.6x    \\ \hline
DGAplus-Oort                                                        & 4.0   & 1.2x    \\ \hline
FedEx                                                           & \textbf{2.3}   & \textbf{2.1x}    \\ \hline
FedAvg                                                              & 4.8   & 1x      \\ \bottomrule
\end{tabular}
\vspace{-3mm}
\end{table}
\vspace{-1mm}

\begin{figure*}[] \centering
  \subfigure[CNN@MNIST.\label{Figure:setKmnist}]
  {\includegraphics[width=0.25\linewidth]{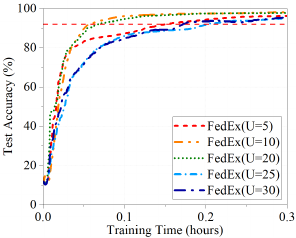}}
  \subfigure[Squeezenet@CIFAR10.\label{Figure:LSTMsen}]
  {\includegraphics[width=0.24\linewidth]{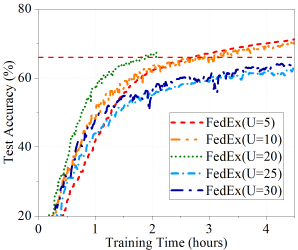}}
  \subfigure[LSTM@Shakespeare.\label{Figure:LSTMsen}]
  {\includegraphics[width=0.24\linewidth]{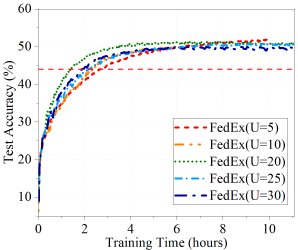}}
  \subfigure[CNN@HAR.\label{Figure:LSTMsen}]
  {\includegraphics[width=0.24\linewidth]{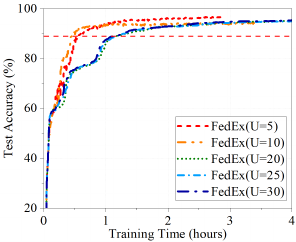}}
  \vspace{-2mm}
 \caption{FedEx performance comparison with different staleness ceilings.} \label{fig:staleness_ceiling}
\vspace{-7mm}
\end{figure*}

\begin{table}\centering
   \caption{Speedup with different trigger thresholds.
   \label{trigger}}
\vspace{-2mm}
\small
\begin{tabular}{c|cc|c}
\toprule
 & \multicolumn{2}{c|}{FedEx} &  \\ \cline{2-3}
FL Task & \multicolumn{1}{c|}{Trigger} & Starting & Speedup \\
 & \multicolumn{1}{c|}{Threshold} & Round &  \\ \midrule
 & \multicolumn{1}{c|}{0.2} & 18 & 1.2x \\ \cline{2-4} 
 & \multicolumn{1}{c|}{0.5} & 156 & 1.8x \\ \cline{2-4} 
Mobileformer@TSRD & \multicolumn{1}{c|}{\textbf{0.7}} & \textbf{207} & \textbf{2.0x} \\ \cline{2-4} 
(Target Acc: 48\%) & \multicolumn{1}{c|}{0.9} & 284 & 1.7x \\ \cline{2-4} 
 & \multicolumn{2}{c|}{FedEx w/o trigger} & 0.9x \\ \cline{2-4} 
 & \multicolumn{2}{c|}{Oort} & 1.6x \\ \cline{2-4} 
 & \multicolumn{2}{c|}{FedAvg} & 1x \\ \midrule
 & \multicolumn{1}{c|}{0.2} & 2 & 1.4x \\ \cline{2-4} 
 & \multicolumn{1}{c|}{0.5} & 14 & 1.7x \\ \cline{2-4} 
CNN@MNIST & \multicolumn{1}{c|}{\textbf{0.7}} & \textbf{39} & \textbf{1.8x} \\ \cline{2-4} 
(Target Acc: 92\%) & \multicolumn{1}{c|}{0.9} & 71 & 1.6x \\ \cline{2-4} 
 & \multicolumn{2}{c|}{FedEx w/o trigger} & 1.4x \\ \cline{2-4} 
 & \multicolumn{2}{c|}{Oort} & 1.5x \\ \cline{2-4} 
 & \multicolumn{2}{c|}{FedAvg} & 1x \\ \midrule
 & \multicolumn{1}{c|}{0.2} & 4 & 1.3x \\ \cline{2-4} 
 & \multicolumn{1}{c|}{0.5} & 22 & 1.7x \\ \cline{2-4} 
LSTM@Shakespeare & \multicolumn{1}{c|}{\textbf{0.7}} & \textbf{52} & \textbf{2.1x} \\ \cline{2-4} 
(Target Acc: 44\%) & \multicolumn{1}{c|}{0.9} & 118 & 1.7x \\ \cline{2-4} 
 & \multicolumn{2}{c|}{FedEx w/o trigger} & 1.3x \\ \cline{2-4} 
 & \multicolumn{2}{c|}{Oort} & 1.6x \\ \cline{2-4} 
 & \multicolumn{2}{c|}{FedAvg} & 1x \\ \bottomrule
\end{tabular}
\vspace{-6mm}
\end{table}
\vspace{-3mm}
\subsection{Advantages of Overlapping Aware PS}
Next, we conduct the ablation study of FedEx's overlapping aware PS utility function, and compare FedEx with DGAplus (random PS), DGAplus-Oort (Oort PS) on the learning task LSTM@Shakespeare. As shown in Fig.~\ref{fig:utility}, FedEx reaches the target accuracy with the smallest latency. Actually, it maintains the highest delay efficiency throughout the entire FL training process. That is because FedEx employs the proposed overlapping aware PS utility function, whose PS jointly considers statistical utility and overlapping latency utility, i.e., the latency reduced by overlapping/continuous computing. Conversely, the random PS in DGAplus considers neither statistical utility nor delay efficiency of FL training, and thus has the worst learning accuracy and latency performance. DGAplus-Oort directly adopts Oort's PS utility function, and ignores the latency reduction benefits of overlapping. Thus, its performance is inferior to FedEx. It can be evidenced by the results in Table~\ref{table:Utility}, where FedEx speeds up FL training to reach the target accuracy by around 1.6x compared with FedAvg, whereas DGAplus-Oort achieves only a $1.2$x acceleration over FedAvg. In addition, similar to DGA, DGAplus has the model drift issue due to overlapping at early FL training stages. Although random PS helps to mitigate it a little bit, the FL training latency in DGAplus is still $1.7$ times longer than that in FedAvg.

\vspace{-3mm}
\subsection{Analysis of FedEx Overlapping Trigger}

We also conduct the ablation study of FedEx overlapping trigger, and compare FedEx having different triggers with FedEx w/o trigger (i.e., FedEx overlapping right after the start of FL training), Oort and FedAvg on three different FL tasks: Mobileformer@TSRD, CNN@MNIST, and LSTM@Shakespeare. We choose trigger threshold $\delta_{CKA}$ = 0.2, 0.5, 0.7 and 0.9, respectively, and present the results in Table~\ref{trigger}. Again, achieving a target accuracy, we let the FL training latency of FedAvg serve as the benchmark (i.e., 1x). We found that setting a too small or too big trigger threshold in FedEx may not have very good performance in latency reduction. If the trigger threshold is too small (e.g., $\delta_{CKA}$ = 0.2), participating mobile devices may start overlapping and continuous computing at very early FL training stages. That may result in model drifts, backfiring FL convergence speed and wasting the computing and memory resources of mobile devices, which verify our analysis in Sec.~\ref{sec:floodtrigger}. The same reason also applies to FedEx w/o trigger. If the trigger is too big (e.g., $\delta_{CKA}$ = 0.9), it may be too late to start overlapping, so that FedEx has missed many opportunities for further latency reduction. From the results in Table~\ref{trigger}, among the chosen FedEx trigger values, $\delta_{CKA}$ = 0.7 yields the best speedup performance, i.e., 2.0x speedup and overlapping starts at Round 207 for MobileFormer@TSRD, 1.8x speedup and overlapping starts at Round 39 for CNN@MNIST, and 2.1x speedup and overlapping starts at Round 82 for LSTM@Shakespeare. Although $\delta_{CKA}$ = 0.7 has good performance, it may not be the optimal overlapping trigger in FedEx. We will leave how to set the optimal trigger for our future studies.

\vspace{-2.5mm}
\subsection{Impacts of Different Staleness Ceilings}
\vspace{-0.5mm}

We also study the impacts of different staleness ceilings on the FL training performance. Here, we keep the classical local iterations at $K=10$, consistent with the other experiments, and vary the upper bounds of the staleness ceiling $U$ to $5$, $10$, $20$, $25$ and $30$ iterations (i.e., $0.5K$, $K$, $2K$, $2.5K$, $3K$, respectively). We conduct experiments on four FL tasks and show the results in Fig.~\ref{fig:staleness_ceiling}.

From the results, we find that the staleness ceilings trade-off double-edges of overlapping/continuous computing (i.e., training speedup and model drift) across various tasks, which validates our analysis in Sec.~\ref{sec:generalization_ceiling}. While overlapping helps to accelerate FL training, it suffers from model drift issues in heterogeneous environments, which may conversely slow down the FL convergence and potentially diminish the overlapping speedup benefits. Specifically, if the value of staleness ceiling is too small, say $0.5 K$, it restricts model drift but limits overlapping acceleration, which results in slow convergence. Meanwhile, if the value of staleness ceiling is too big, e.g., $30$ ($3K$), it makes model drift dominant, which also results in slow convergence. Aside from two extreme ends, an appropriate selection of staleness ceilings will help overlapping speedup benefits outweigh the convergence slowdown due to model drift, and effectively reduce the FL training latency. Besides, the good choice of staleness ceiling is learning task specific. For example, as shown in Fig.~\ref{fig:staleness_ceiling}, for the CNN@MNIST and CNN@HAR, $U = 10$ (around $K$) has the best performance. For Squeezenet@CIFAR10 and LSTM@Shakespeare, a relatively large staleness ceiling (i.e., $U = 20$, around $2K$) would be favored.
\begin{figure} \centering
  \subfigure[CNN@HAR.\label{Figure:CNNsen}]
  {\includegraphics[width=0.49\linewidth]{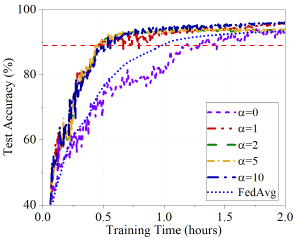}}
  \subfigure[LSTM@Shakespeare.\label{Figure:LSTMsen}]
  {\includegraphics[width=0.48\linewidth]{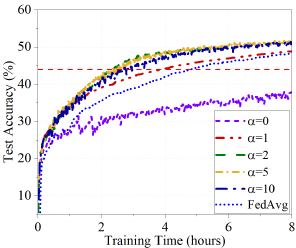}}
\vspace{-2mm}
 \caption{FedEx sensitivity with different $\alpha$ values.} \label{fig:alpha}
\vspace{-4mm}
\end{figure}

\vspace{-2mm}
\subsection{FedEx Sensitivity Analysis}

FedEx uses the scaling coefficient $\alpha$ to balance overlapping latency utility and statistical utility, where it adaptively prioritizes high overlapping latency utility participating devices, and penalizes the utility of stragglers in PS. Figure~\ref{fig:overall} and Table~\ref{table:comparisonSOTA} show that FedEx ($\alpha = 2$) outperforms its counterparts, and Fig.~\ref{fig:alpha} shows that FedEx achieves similar performance across all non-zero $\alpha$ values ($\alpha = 1, 2, 5$ or $10$). The results validate that FedEx can orchestrate its components to automatically navigate for the desired FL performance and training performance is not that sensitive to $\alpha$ values. When setting $\alpha=0$, the FL server solely relies on statistical utility to select participating devices, which overlooks the time-saving benefits brought by overlapping yet amplifies its impacts of model drifts, so it has even worse performance than FedAvg.


\begin{figure} \centering
  \subfigure[CNN@MNIST.\label{Figure:setKmnist}]
  {\includegraphics[width=0.49\linewidth]{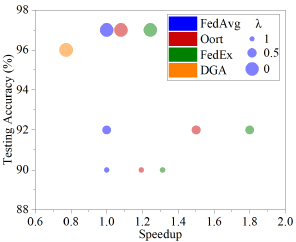}}
  \subfigure[Squeezenet@CIFAR10.\label{Figure:SqC10dataheter}]
  {\includegraphics[width=0.48\linewidth]{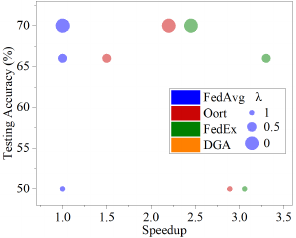}}
\vspace{-2mm}
 \caption{Impacts of data heterogeneity.} \label{fig:data}
\vspace{-5mm}
\end{figure}

\vspace{-3mm}
\subsection{Impacts of Data Heterogeneity}
We further evaluated the performance of FedEx under various non-i.i.d. levels of training data samples. Figure~\ref{fig:data} shows the testing accuracy and corresponding speedups under non-i.i.d. levels of $\lambda = 0$, $0.5$ and $1$, respectively, where the latency of FedAvg is the benchmark. Compared with other designs, FedEx always has the most speedup to reach the target accuracy, especially at $\lambda=0.5$. In particular, for the i.i.d. data distribution case (i.e., $\lambda=0$) on CNN@MNIST, we find that FedEx is better than Oort or DGA in terms of the training latency to reach the target accuracy. The potential reason is that the computing and wireless heterogeneity still exist among selected mobile devices, even though the data is i.i.d.. FedEx considers the latency reduction both by overlapping and by addressing straggler issues per round, and works well in such heterogeneous environments. Besides, for the non-i.i.d. cases on CNN@MNIST, DGA has the worst performance (below 88\% lower bound of testing accuracy shown in Fig.~\ref{fig:data}) or even fail FL convergence because of deficiencies discussed in Sec.~\ref{sec:defDGA}. 
As for FedEx, Oort and FedAvg, we observe the same performance trend across various $\lambda$ values for Squeezenet@CIFAR10, as shown in Fig.~\ref{Figure:SqC10dataheter}. For DGA, its performance is not included in Fig.~\ref{Figure:SqC10dataheter}, since DGA's testing accuracy remains below $50\%$ across different $\lambda$ values in heterogeneous environments. In particular, for the i.i.d. case, the convergence speed of DGA is even slower than that of FedAvg. This is because DGA requires all devices to participate in each round, which involves the straggler every time in the heterogeneous FL setting. FedAvg mitigates this issue by randomly sampling devices for participation in each FL training round.

\begin{table}\centering
   \caption{Impact of system heterogeneity. (H: High-end, Xiaomi 12S; M: Middle-end, Honor 70; L: Low-end, NVIDIA TX2)
   \label{table:diffHeteo}}
   \vspace{-3mm}
\small
\vspace{1mm}
\begin{tabular}{c|ccc}
\toprule
 & \multicolumn{3}{c}{OL of different type of system heterogeneity (h)} \\ \hline
Methods & \multicolumn{1}{c|}{70H/20M/10L} & \multicolumn{1}{c|}{30H/30M/40L} & 20H/30M/50L \\ \hline
FedAvg & \multicolumn{1}{c|}{5.15} & \multicolumn{1}{c|}{5.35} & 5.41 \\ \hline
Oort & \multicolumn{1}{c|}{4.58} & \multicolumn{1}{c|}{4.64} & 4.89 \\ \hline
DGA & \multicolumn{3}{c}{Cannot Reach Target Accuracy} \\ \hline
FedEx & \multicolumn{1}{c|}{1.62} & \multicolumn{1}{c|}{2.38} & 3.74 \\ \bottomrule
\end{tabular}
\vspace{-5mm}
\end{table}

\vspace{-3mm}
\subsection{Impacts of System Heterogeneity}

We also explore the performance of FedEx under different types of heterogeneity, as shown in Table \ref{table:diffHeteo}. We select three devices with significantly different computational and communication capabilities for emulation experiments: Xiaomi 12S as High-end (H), Honor 70 as Middle-end (M), and NVIDIA TX2 as Low-end (L). Keeping the total number of devices at 100, we set up three device distribution schemes: 70H/20M/10L (indicating 70 High-end, 20 Middle-end, and 10 Low-end devices); 30H/30M/40L; and 20H/30M/50L. We compare the overall latency (OL) to reach the target accuracy in FL training under these three settings using LSTM@Shakespe-are. The experimental results show that as the number of Low-end devices increases, the OL increases for all methods. In all types of system heterogeneity, FedEx significantly outperforms Oort and FedAvg. Since the computation latency of the Low-end TX2 is much greater than its communication latency, the benefit from overlap training decreases as the proportion of such devices increases, resulting in a reduced improvement from FedEx.

\vspace{-3mm}
\section{Conclusion}
In this paper, we developed FedEx to enable overlapping in heterogeneous environments and further reduce FL training latency by the proposed overlapping aware PS, staleness ceilings and triggers. Observing (i) the model staleness, model drifts, memory cost and straggler issues inherent in the original overlapping-based FL protocol under heterogeneous data, device, and wireless environments, and (ii) the incompatibility of existing PS policies for overlapping based FL, we redesigned the overlapping computation procedure in FedEx with staleness ceiling and devised a new overlapping aware PS solution to alleviate the straggler effects and harness the benefits of continuous local computation. Addressing model drift concerns, FedEx also integrates a threshold-based method to trigger overlapping training to avoid the waste of devices' resources. Through extensive experiments, we demonstrated the superior performance of FedEx in achieving training speedup with limited memory costs. FedEx builds the framework for overlapping based FL training over heterogeneous devices and its further improvement/integration with other pioneering designs dealing with system heterogeneity (i.e., HeteroFL~\cite{diao2021heterofl}, FedRolex~\cite{FedRolexNIPS22}, etc.).

\bibliographystyle{IEEEtran}
\bibliography{citations_TMC}



\begin{IEEEbiography}[{\includegraphics[width=1in,height=1.25in,clip,keepaspectratio]{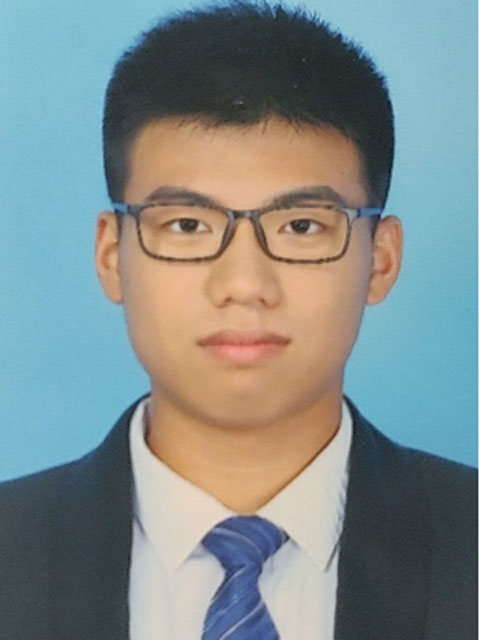}}]{Jiaxiang Geng}
(S'22) received his B.S. degree in Information Engineering from Beijing University of Posts and Telecommunications (BUPT) in 2022. He is currently working toward the M.S. degree in Information and Communication Engineering with the School of Information and Communication Engineering, BUPT. His research interests include Federated Learning, Mobile Edge Computing and Deep Learning.
\end{IEEEbiography}

\begin{IEEEbiography}[{\includegraphics[width=1in,height=1.25in,clip,keepaspectratio]{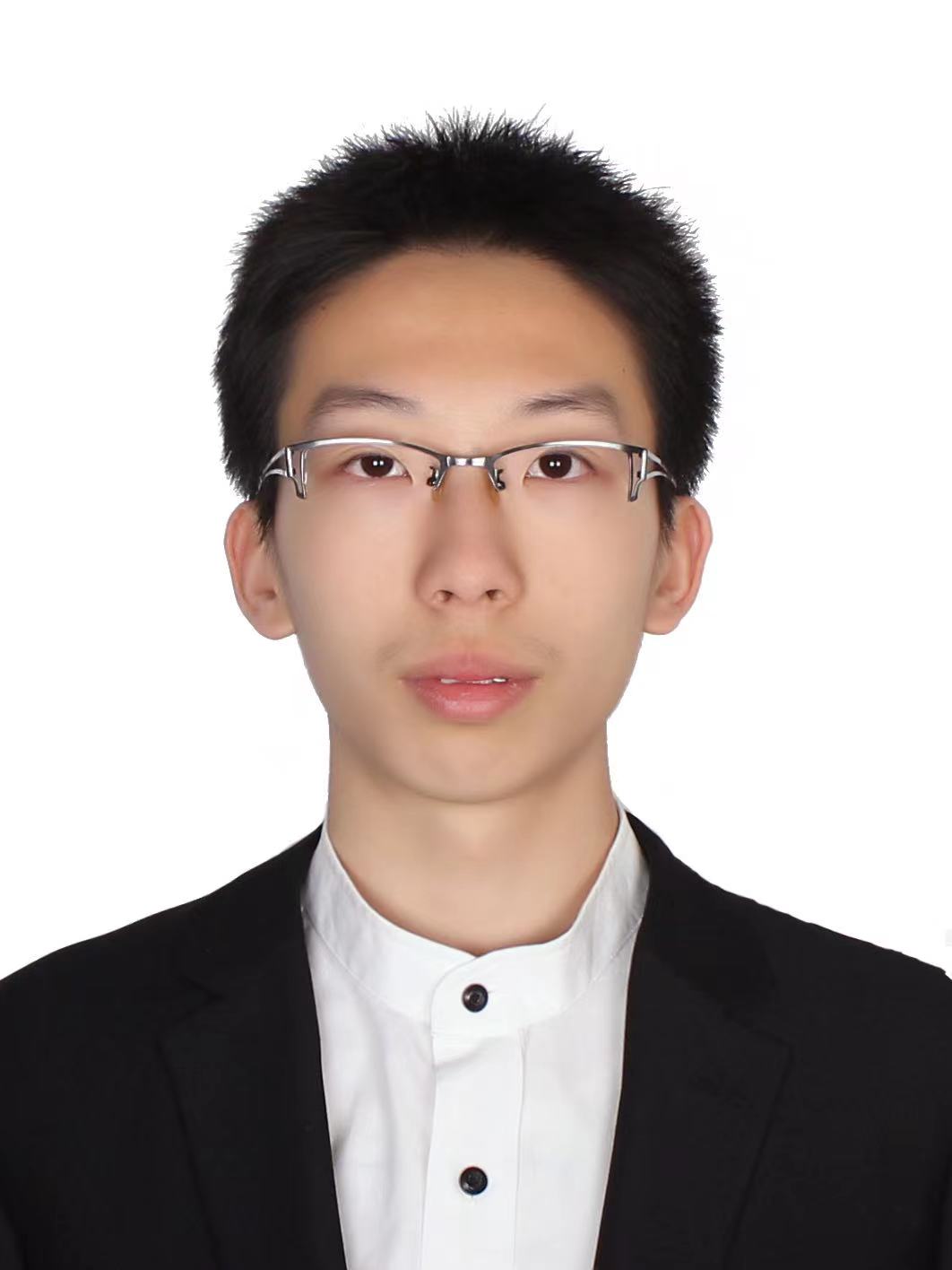}}]{Boyu Li} received his B.S. degree in Communication Engineering from Beijing University of Posts and Telecommunications (BUPT) in 2023. He is currently working toward the M.S. degree in Information and Communication Engineering with the School of Information and Communication Engineering, BUPT. His research interests include Federated Learning and Edge Computing.
\end{IEEEbiography}

\begin{IEEEbiography}[{\includegraphics[width=1in,height=1.25in,clip,keepaspectratio]{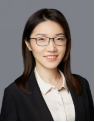}}]{Xiaoqi Qin}
(S’13-M’16-SM’23) received her B.S., M.S., and Ph.D. degrees from Electrical and Computer Engineering with Virginia Tech. She is currently an Associate Professor of School of Information and Communication Engineering with Beijing University of Posts and Telecommunication (BUPT). Her research mainly focuses on task-oriented machine-type communications and networked intelligence. She has published more than 80 journal and conference papers, one book, and holds 21 patents on these areas. She was a Distinguished Young Investigator of China Frontiers of Engineering. She has received the Best Paper Awards at IEEE GLOBECOM’23 and WCSP’23. She was a recipient of first Prize of Science and Tech. Progress Award by Chongqing Municipal People's Government, and first Prize of Tech. Invention Award by China Institute of Communications. She has served as the symposium lead chair, the publicity co-chair, and member of the Technical Program Committee for several international conferences. 
\end{IEEEbiography}

\begin{IEEEbiography}[{\includegraphics[width=1in,height=1.25in,clip,keepaspectratio]{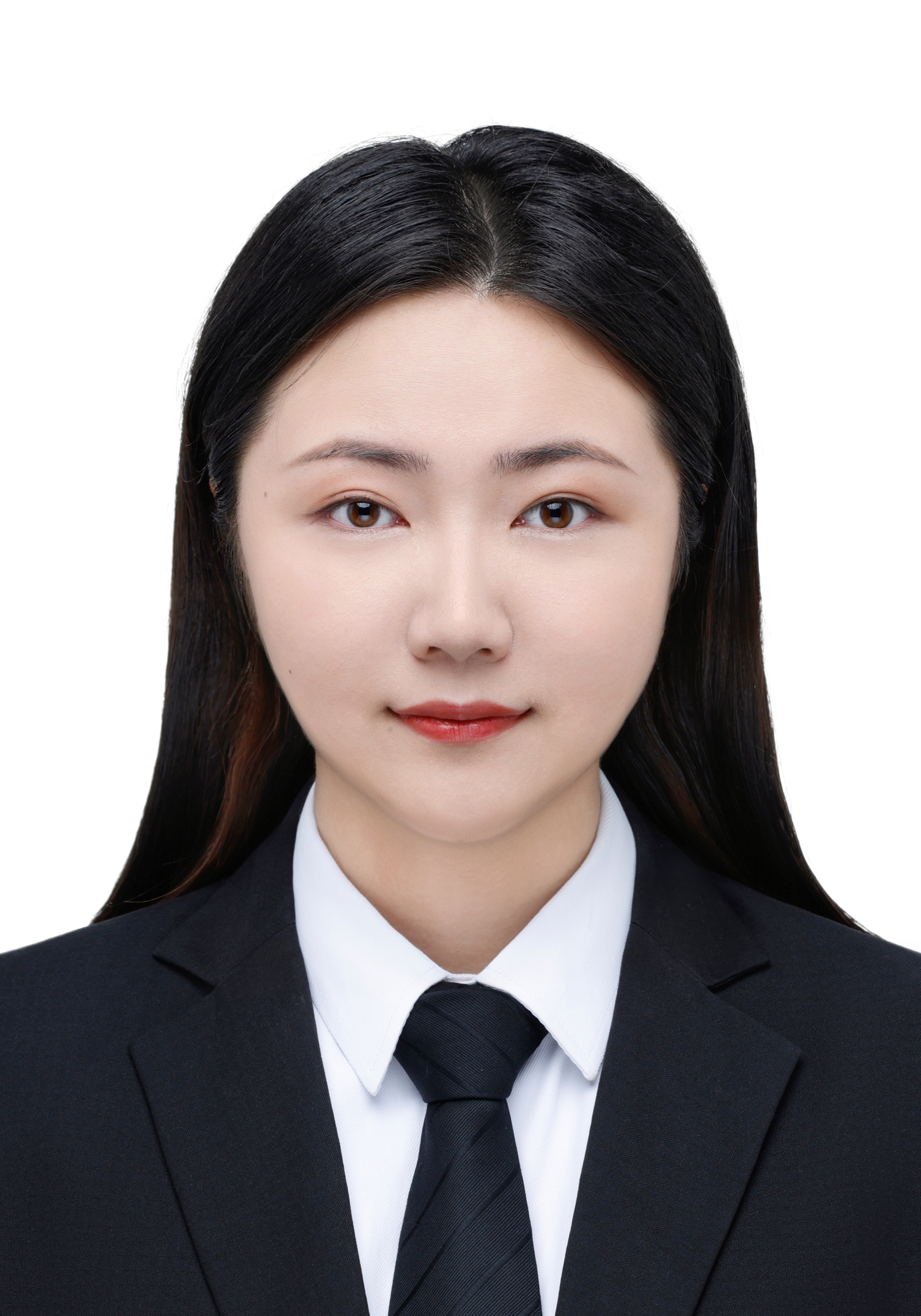}}]{Yixuan Li}
(S’21) received the Ph.D. degree in School of Information and Communication Engineering at Beijing University of Posts and Telecommunications (BUPT) in 2024. She is now a lecturer at the College of Electronic Information Engineering at Taiyuan University of Technology, China. Her research interests include wireless for AI, mobile edge computing, and energy efficient communication systems.
\end{IEEEbiography}

\begin{IEEEbiography}[{\includegraphics[width=1in,height=1.25in,clip,keepaspectratio]{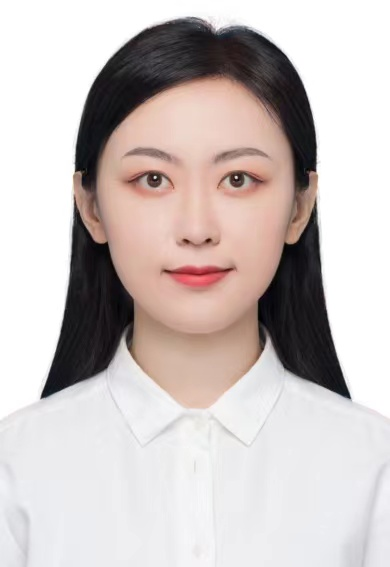}}]{Liang Li}
Liang Li (S’19-M’21) received the Ph.D. degree in the School of Telecommunications Engineering at Xidian University, China, in 2021. She was a post-doctoral faculty member with the School of Computer Science (National Pilot Software Engineering School), Beijing University of Posts and Telecommunications, from 2021 to 2023. Since 2023, she has been with Pengcheng Laboratory, China, where she is currently an assistant researcher with the Department of Advanced Interdisciplinary Research. She was also a visiting Ph.D. student with the Department of Electrical and
Computer Engineering, University of Houston, Houston, TX, USA, from 2018 to 2020. Her research interests include edge intelligence, federated learning, edge computing and caching, data-driven robust optimization, and differential privacy.
\end{IEEEbiography}

\begin{IEEEbiography}[{\includegraphics[width=1in,height=1.25in,clip,keepaspectratio]{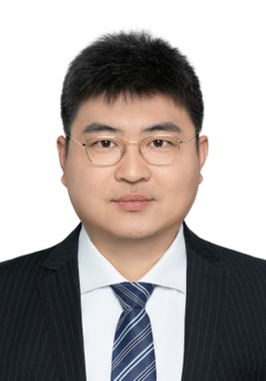}}]{Yanzhao Hou}
received his Ph.D. degree from the Beijing University of Posts and Telecommunications (BUPT), Beijing, China, in 2014. He is currently an  Associate Professor with the National Engineering Research Center for Mobile Network Technologies, BUPT. His current research interests include federated learning, software defined radio, terahertz communications and trial systems. He received the Best Demo Award in IEEE APCC2018.
\end{IEEEbiography}

\begin{IEEEbiography}[{\includegraphics[width=1in,height=1.25in,clip,keepaspectratio]{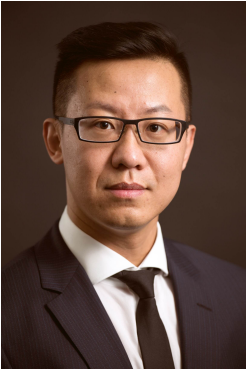}}]{Miao Pan}
(S’07-M’12-SM’18) Dr. Miao Pan is a Full Professor in the Department of Electrical and Computer Engineering at University of Houston. He was a recipient of NSF CAREER Award in 2014. Dr. Pan received his Ph.D. degree in Electrical and Computer Engineering from University of Florida in August 2012. Dr. Pan's research interests include wireless for AI, mobile AI systems, deep learning privacy, new biometric based authentication, quantum computing privacy, cybersecurity, wireless networks, and underwater IoT and networks.
Dr. Pan is an Associate Editor for ACM Computing Surveys, IEEE Journal of Oceanic Engineering, IEEE Open Journal of Vehicular Technology and IEEE Internet of Things (IoT) Journal (Area 5: Artificial Intelligence for IoT), and used to be an Associate Editor for IEEE Internet of Things (IoT) Journal (Area 4: Services, Applications, and Other Topics for IoT) from 2015 to 2018. Dr. Pan is a member of AAAI, a member of ACM, and a senior member of IEEE and IEEE Communications Society.
\end{IEEEbiography}

\newpage

\hyphenation{op-tical net-works semi-conduc-tor IEEE-Xplore}
\def\BibTeX{{\rm B\kern-.05em{\sc i\kern-.025em b}\kern-.08em
    T\kern-.1667em\lower.7ex\hbox{E}\kern-.125emX}}

\appendices
\section{FedEx Algorithm}
\begin{algorithm}[H]
\small
  \caption{FedEx Procedure}\label{Alg:FedEx}
  \begin{algorithmic}[1]
    \State \textbf{Input:} Mobile device set $\mathcal{N}$ with size $N$, participant size $P$, scaling coefficient $\alpha$, classical local iterations $K$, learning rate $\eta$.
    \State $\mathcal P^r$ is the participants in round $r$, $m_n^r$ is the local updates record, $\overline{m^r}$ is the global updates record, $w_n^{r,k}$ is the local model and $g_n^{r,k}$ is the local gradient.
    \State At the edge server, initialize participants $\mathcal P \leftarrow \emptyset$, training round $r \leftarrow 0$, utility value of each mobile device $Util \leftarrow \emptyset$.
    \State Oort is conducted until activated by FedEx overlapping trigger Eqn.(4).
    \State For the $r$-th round FL training, where overlapping is activated: 
      \State \underline{\textbf{On Edge Server:}}
      \State Broadcast a ``finished signal'' to all devices when the last device selected in the previous round finishes.
      \State Receive the utility parameters from mobile devices.
      \State Aggregate the received local updates $\overline{m^{r-1}} \leftarrow \frac{1}{P}\sum_{n \in \mathcal P^{r-1}}m_n^{r-1}$ as global updates.
      \State Calculate $Util(n,r)$ according to Eqn.(3).
      \State $\mathcal P^r =$ RankingDevice($Util(n,r)$).
      \State Broadcast the device selection decision $\mathbb{V}(n,r)$ and global updates $m^{r-1}$ to selected devices.
      \State \underline{\textbf{On Mobile Devices:}}
      \For{$n \in \mathcal{N}$ in parallel}
          \If{$\mathbb{V}(n,r) = 1$}
                \State $w_n^{r,0} \leftarrow w_n^{r-1,S_n^{r-1}}+m_n^{r-1}-\overline{m^{r-1}}$
              \State Calculate $K-S_n^{r-1}$.
              \State $u_n^{r} \leftarrow 0$.
              \State Perform local computing for $K-S_n^{r-1}$ iterations to get $w_n^{r,K-S_n^{r-1}} \leftarrow w_n^{r,0}-\eta \sum_{k=1}^{K-S_n^{r-1}} g_n^{r,k}$. 
              \State Record local updates $m_n^r \leftarrow \eta \sum_{k=1}^{K-S_n^{r-1}} g_n^{r,k}$.
              \State Transmit the local updates $m_n^r$ to FL Sever while proceeding with continuous computing.
              \State Continuous computing stops when receiving a "finished signal" from the server or reaching $K$ overlapping iterations.
              \State Record the overlapping computing iterations $S_n^r$ and the local model $w_n^{r,S_n^r} \leftarrow w_n^{r,K-S_n^{r-1}} - \eta \sum_{s=1}^{S_n^r}g_n^{r,s}$
          \Else
              \State $u_n^{r} \leftarrow u_n^{r-1}+1$ \Comment{Non-participating rounds}
              \State $S_n^{r} \leftarrow S_n^{r-1}$ \Comment{Overlapping computing iterations}
              \State $m_n^{r} \leftarrow m_n^{r-1}$ \Comment{Local Updates Record}
          \EndIf 
              \State {Calculate/estimate $|B_n^r|$, $Loss$, $t_n^r$, $D_n$, $R_n^r$}
              \State Send $|B_n^r|$, $Loss$, $t_n^r$, $D_n$, $R_n^r$, $S_n^{r}$, $K$ to FL server
      \EndFor
      \State \textbf{Return} $w^r=\frac{1}{N}\sum_{n=1}^N w_n^{r,K-S_n^{r-1}}$

  \end{algorithmic}

\end{algorithm}

\section{Theoretical Analysis}
\numberwithin{equation}{section}
\subsection{Preliminaries}
Jensen's inequalities are frequently used throughout our theoretical analysis. We will use it without further explanation. From Jensen's inequaility, for any $z_m \in \mathbb{R}^d$, $m \in {1,2,...,M}$, we have
\vspace{-2mm}
\begin{equation}
\small
 \begin{aligned}
 ||\frac{1}{M}\sum_{m=1}^M z_m ||^2 \le \frac{1}{M} \sum_{m=1}^M ||z_m||^2
 \end{aligned}
 \vspace{-2mm}
\end{equation}
\vspace{-2mm}
which directly gives
\begin{equation}
\small
 \begin{aligned}
 ||\sum_{m=1}^M z_m ||^2 \le M\sum_{m=1}^M ||z_m||^2
 \end{aligned}
 \vspace{-2mm}
\end{equation}
Additionally, we standardize the symbols used throughout this theoretical analysis as shown in Appendix Table 1.
\begin{table}[ht]
\renewcommand{\tablename}{Appendix Table}
    \label{param}
    \small
    \caption{Symbol Definition}
    \vspace{-2mm}
    \centering
    \begin{tabular}{|l|p{0.6\linewidth}|}
        \hline
        \textbf{Symbol} & \textbf{Description} \\
        \hline
        \( N \) & Total number of mobile devices \\
        \hline
        \( P \) & Number of mobile devices participating in each training round \\
        \hline
        \( K \) & Number of classical local computation iterations \\
        \hline
        \( S_n^r \) & Overlapping iterations of client \( n \) in the \( r \)-th round \\
        \hline
        \( w_n^{r,k} \) & Local model of client \( n \) in the \( r \)-th round after k iterations of local update\\
        \hline
        \( w^r \) & Global model \\
        \hline
        \( F(\cdot) \) & Global loss function \\
        \hline
        \( F_n(\cdot) \) & Local loss function of client \( n \) \\
        \hline
        \( g_n^{r,k} \) & Gradient value of client \( n \) in the \( k \)-th gradient computation of the \( r \)-th round \\
        \hline
    \end{tabular}
\vspace{-4mm}
\end{table}

For theoretical derivations, we assume that the device selection method by FedEx results in \(\mathbf{q} = \{q_1, q_2, ..., q_N\}\), representing the participation probability of each device, where $\sum_{n=1}^N q_n=P$. The global model $w^r$ for each round can be represented by 
\vspace{-4mm}
\begin{equation}
\small
\label{aggregation}
 \begin{aligned}
 w^r=\frac{1}{N}\sum_{n=1}^N w_n^{r,K-S_n^{r-1}}
 \end{aligned}
\vspace{-2mm}
\end{equation}

In this theoretical analysis, we focus on the overlapping phase of FedEx with a staleness ceiling \( K \).

\subsection{A step by step walk through FedEx}

Here, we provide a step-by-step walk through of FedEx to demonstrate its effectiveness under conditions of clients' partial participation and system heterogeneity. We set the very first round of overlap training as \( r = 1 \) and the initial model as $w_0$. For device \( n \) in \( r = 1 \), with a probability \( q_n \) of participating, solely local updates are performed; hence, after completing \( K \) iterations of classical local computing, the local model can be represented as:
\vspace{-4mm}
\begin{equation}
\small
 \begin{aligned}
\mathbb{E}[w_n^{1,K}]=w_0-\eta q_n \sum_{k=1}^K g_n^{1,k}
 \end{aligned}
\vspace{-6mm}
\end{equation} \\

At this time, client \( n \) sends the local updates (accumulated gradient) \( m_n^1 = \eta\sum_{k=1}^K g_n^{1,k} \) to the server. While uploading the local updates, client \( n \) continues executing overlapping local computing iterations. Continuous computing stops when receiving a "finished signal" from the server or reaching \( K \) overlapping iterations. The actual overlapping iterations performed by client \( n \) are recorded as \( S_n^1 \), and the local model can be represented as:
\vspace{-4mm}
\begin{equation}
\small
 \begin{aligned}
\mathbb{E}[w_n^{1,K+S_n^1}]=w_0-\eta q_n \sum_{k=1}^K g_n^{1,k}-\eta q_n \sum_{s=1}^{S_n^1}g_s^{1,s}
 \end{aligned}
\vspace{-6mm}
\end{equation} \\

The server aggregates the local updates from all participating clients to obtain the global updates. Based on the probability assumptions of client selection (partial participation) made in Section C.1, the expected value of the global updates can be represented as:
\vspace{-4mm}
\begin{equation}
\small
 \begin{aligned}
\mathbb{E}[\overline{m^1}]=\frac{1}{N} \sum_{n=1}^N q_n m_n^1=\frac{\eta}{N}\sum_{n=1}^N q_n \sum_{k=1}^K g_n^{1,k}
 \end{aligned}
\vspace{-6mm}
\end{equation} \\

According to Eqn. \eqref{aggregation}, the global model here can be represented as:
\vspace{-4mm}
\begin{equation}
\small
 \begin{aligned}
\mathbb{E}[w^1]=\frac{1}{N}\sum_{n=1}^N  w_n^{1,K}
=w_0-\frac{\eta}{N}\sum_{n=1}^{N}q_n \sum_{k=1}^K g_n^{1,k}
 \end{aligned}
\vspace{-2mm}
\end{equation} \\
The server transmits the global updates to each client. Client \( n \) has a probability \( q_n \) of participating in the next round. If it participates, it first performs global updates correction. The local model after correction is represented as:
\begin{equation}
\small
 \begin{aligned}
\mathbb{E}[w_n^{2,0}]&=\mathbb{E}[w_n^{1,K+S_n^1}]+q_n(m_n^1-\overline{m^1}) \\
&=w_0- \eta q_n\sum_{s=1}^{S_n^1}g_n^{1,s}- q_n \frac{\eta}{N}\sum_{n'=1}^N q_n' \sum_{k=1}^K g_n^{1,k}
 \end{aligned}
 \vspace{-6mm}
\end{equation} \\

Then, it executes \( K - S_n^1 \) iterations of classical local computing, and the resulting local model can be represented as:
\vspace{-4mm}
\begin{equation}
\small
 \begin{aligned}
\mathbb{E}[w_n^{2,K-S_n^1}]&=\mathbb{E}[w_n^{2,0}]-\eta q_n \sum_{k=1}^{K-S_n^1} g_n^{2,k} \\
&=w_0- q_n \frac{\eta}{N}\sum_{n'=1}^N q_{n'} \sum_{k=1}^K g_n^{1,k}-\eta q_n \sum_{k=1}^{K} g_n^{2,k}
 \end{aligned}
\vspace{-6mm}
\end{equation} \\

After aggregation by the server, the expected global model for \( r = 2 \) can be represented as:
\begin{equation}
\small
 \begin{aligned}
\mathbb{E}[w^2]&=\frac{1}{N}\sum_{n=1}^N w_n^{2,K-S_n^1} \\
&=w_0-\frac{P\eta}{N^2}\sum_{n=1}^N q_n \sum_{k=1}^K g_n^{1,k}-\frac{\eta}{N}\sum_{n=1}^N q_n\sum_{k=1}^K g_n^{2,k}
 \end{aligned}
\vspace{-6mm}
\end{equation} \\

Therefore, the gradient difference in the global model between \( r=1 \) and \( r=2 \) is:
\begin{equation}
\small
 \begin{aligned}
\mathbb{E}[w^2-w^1]=-\frac{\eta}{N}\sum_{n=1}^N q_n\sum_{k=1}^K g_n^{2,k}-(\frac{P}{N}-1)\frac{\eta}{N}\sum_{n=1}^N q_n \sum_{k=1}^K g_n^{1,k}
 \end{aligned}
\end{equation}

When \( P=N \), meaning all clients participate, the second term is \( 0 \), aligning with the gradient update of the classical FL global model. Moreover, this relationship can be extended to any two adjacent rounds \( r \) and \( r+1 \), expressed as follows:
\begin{equation}
\small
\label{diff}
 \begin{aligned}
\mathbb{E}[w^r-w^{r-1}]=-\frac{\eta}{N}\sum_{n=1}^N q_n\sum_{k=1}^K g_n^{r,k}
 \end{aligned}
\end{equation} 

This explains that FedEx achieves the same convergence as FedAvg when all devices participate in training, regardless of system heterogeneity, thanks to the staleness ceiling we proposed. We will explain in Section C.5 why DGA fails on heterogeneous devices. However, as mentioned in our main text, involving all devices can lead to straggler issues, meaning that involving all devices does not necessarily maximize system efficiency. Therefore, we introduced device selection, where only some devices participate. 

When \( P < N \), indicating partial device participation in training, the second term is non-zero. This reflects that overlap training with partial device participation introduces a certain amount of model drift in the global model. Fortunately, this redundant term gradually diminishes as rounds progress, ensuring it does not adversely affect model convergence. 

Although model drift under partial device participation may temporarily affect the global model in a round, potentially increasing (or decreasing, when the redundant term aligns with the normal gradient direction) the number of rounds required for convergence, careful design of device selection in FedEx significantly enhances system efficiency.

\vspace{-2mm}
\subsection{Assumptions}
To start the convergence analysis, we assume that the objective function is L-smooth. \\
\textbf{Assumption 1} (L-smoothness). The difference with L-Lipschitz gradient satisfies
\begin{equation}
\small
 \begin{aligned}
||\nabla F_n(x)-\nabla F_n(y)|| \le L ||x-y|| \ \ \forall x,y \in \mathbb{R}^d
 \end{aligned}
 \vspace{-4mm}
\end{equation} \\
\textbf{Assumption 2} (Bounded gradients and variances). We already have $\mathbb{E}[g_n(w)]=\nabla F_n(w)$. We assume that the unbiased gradients has bounded second moment and variance:
\begin{equation}
\small
 \begin{aligned}
\mathbb{E}[||g_n(w)||^2]\le G^2, \ \ and \ \ \mathbb{E}[||g_n(w)-\nabla F_n(w)||^2] \le \delta^2 \ \ \forall w,n.
 \end{aligned}
\vspace{-4mm}
\end{equation} \\
Based on Assumption 2, we can derive the following Bounded Variation:\\
\textbf{Lemma 1} (Bounded Variation). The difference between the n-th client and the average parameter is uniformly bounded:
\begin{equation}
\small
 \begin{aligned}
\mathbb{E}[||w_n^{r,k}-w^r||^2] \le 4 \eta^2 K^2 G^2
 \end{aligned}
 \vspace{-1mm}
\end{equation} 

\begin{IEEEproof}
If client \( n \) participates in round \( r \), its local gradients update before being corrected by the global updates will satisfy the following condition:
\begin{equation}
\small
 \begin{aligned}
w_n^{r,k} = w_n^{r,0} - \eta \sum_{k=1}^{K-S_n^{r-1}} g_n^{r,k} - \eta \sum_{s=1}^{S_n^r} g_n^{r,s}
 \end{aligned}
 \vspace{-2mm}
\end{equation} \\
Since \( K-S_n^{r-1} \le K \) and \( S_n^r \le K \), the number of local gradients not aligned is at most \( 2K \). This means the parameter \( w_n^{r,k} \) and \( w^r \) differ by a maximum of \( 2K \) local gradients throughout the entire process.
\end{IEEEproof}

\subsection{Convergence Analysis of FedEx}
\textbf{Theorem 1} Let \( F^* = \min_w F(w) \). When \( \eta < \frac{N}{PKL} \), for \( R \) rounds, the global model sequence \(\{w^r\}\) yielded by FedEx satisfies:
\begin{equation}
\small
 \begin{aligned}
\frac{1}{R}\sum_{r=1}^R\mathbb{E}[||\nabla F(w^r)||^2] \le &\frac{2N}{\eta PKR} (\mathbb{E}[F(w^0)]-F^*)+\frac{L\eta\delta^2}{P} \\
&+\frac{4N^2L^2G^2\eta^2K^2}{P^2}
 \end{aligned}
\end{equation} 

\begin{IEEEproof}
By the L-smoothness in Assumption 1, we have
\begin{equation}
\small
\vspace{-1mm}
 \begin{aligned}
\mathbb{E}[F(w^r)]\le &\mathbb{E}[F(w^{r-1})]+\mathbb{E}[<\nabla F(w^r),w^{r}-w^{r-1}>] \\
&+\frac{L}{2}\mathbb{E}[||w^{r}-w^{r-1}||^2]
 \end{aligned}
\vspace{-2mm}
\end{equation} \\
We use Jensen's inequality to simplify the third term, resulting in:
\begin{equation}
\small
\vspace{-1mm}
 \begin{aligned}
&\mathbb{E}[||w^r\!-\!w^{r-1}||^2]=\eta^2\mathbb{E}[||\frac{1}{N}\sum_{n=1}^N q_n \sum_{k=1}^K g_n^{r,k} ||^2] \\
&\le \!\frac{\eta^2 K}{N}\!\mathbb{E}\![||g_n^{r,k}\!-\!\nabla F_n(w_n^{r,k})||^2]\!+\!\eta^2 \!\mathbb{E}\![\!||\!\frac{1}{N}\!\sum_{n=1}^N\!q_n\!\sum_{k=1}^K \nabla F_n(w_n^{r,k})||^2] \\
&\le \frac{\eta^2 K \delta^2}{N}+\eta^2 \mathbb{E}[||\frac{1}{N}\sum_{n=1}^Nq_n\sum_{k=1}^K \nabla F_n(w_n^{r,k})||^2]
 \end{aligned}
\end{equation} \\

\begin{table*}[]
\centering
\renewcommand{\tablename}{Appendix Table}
\label{heterogeneity_app}
\small
\caption{Additional Experiments on Impact of System Heterogeneity (LSTM@Shakespeare)}
\begin{tabular}{c|ccccc|cccc}
\hline
 & \multicolumn{5}{c|}{\multirow{2}{*}{Number of devices (Total number of devices: 100)}} & \multicolumn{4}{c}{Methods} \\ \cline{7-10} 
 & \multicolumn{5}{c|}{} & FedAvg & Oort & DGA & FedEx \\ \cline{2-10} 
 & Xiaomi12S & Xavier & Honor70 & HonorPlay6T & TX2 & \multicolumn{4}{c}{Overall Latency (h)} \\ \hline
\multirow{6}{*}{\begin{tabular}[c]{@{}c@{}}Different\\ types\\ of\\ system\\ heterogeneity\end{tabular}} & 60 & 10 & 10 & 10 & 10 & 5.14 & 4.41 & \multirow{6}{*}{\begin{tabular}[c]{@{}c@{}}Cannot\\ reach\\ target\\ acc\end{tabular}} & \textbf{2.86} \\
 & 10 & 60 & 10 & 10 & 10 & 5.20 & 4.68 &  & \textbf{3.45} \\
 & 10 & 10 & 60 & 10 & 10 & 5.22 & 4.75 &  & \textbf{4.34} \\
 & 10 & 10 & 10 & 60 & 10 & 5.27 & 4.88 &  & \textbf{4.72} \\
 & 10 & 10 & 10 & 10 & 60 & 5.43 & 4.97 &  & \textbf{4.81} \\
 & 20 & 20 & 20 & 20 & 20 & 5.21 & 4.70 &  & \textbf{4.08} \\ \hline
\end{tabular}
\end{table*}

Moreover,
\begin{equation}
\small
\vspace{-1mm}
 \begin{aligned}
&\mathbb{E}[<\!\nabla F(w^r),w^r\!-\!w^{r-1}\!>]\!=\!E[<\!\nabla F(w^r)\!,-\frac{\eta}{N}\sum_{n=1}^N \!q_n\! \sum_{k=1}^K g_n^{r,k}\!>] \\
&=-\frac{\eta N}{PK} \mathbb{E}[<\frac{PK}{N}\nabla F(w^r),\frac{1}{N}\sum_{n=1}^N q_n \sum_{k=1}^K \nabla F_n(w_n^{r,k})>] \\
&=\frac{\eta N}{2PK} \mathbb{E}[||\frac{PK}{N}\nabla F(w^r)-\frac{1}{N}\sum_{n=1}^N q_n \sum_{k=1}^K \nabla F_n(w_n^{r,k})||^2]  \\
&-\frac{\eta PK}{2N}\mathbb{E}[||\nabla F(w^r)||^2]-\frac{\eta N}{2PK}\mathbb{E}[||\frac{1}{N}\sum_{n=1}^N q_n \sum_{k=1}^K \nabla F_n(w_n^{r,k})||^2]
 \end{aligned}
\vspace{-2mm}
\end{equation} \\
With Lemma 1, the first term can be upper bounded by:
\begin{equation}
\small
\vspace{-1mm}
 \begin{aligned}
&\mathbb{E}[||\frac{PK}{N}\nabla F(w^r)-\frac{1}{N} \sum_{n=1}^N q_n \sum_{k=1}^K \nabla F_n(w_n^{r,k})||^2] \\
&=\mathbb{E}[||\frac{1}{N}\sum_{n=1}^N q_n \sum_{k=1}^K \nabla F(w^r)-\frac{1}{N}\sum_{n=1}^N q_n \sum_{k=1}^K \nabla F_n(w_n^{r,k})||^2] \\
&=\mathbb{E}[||\frac{1}{N} \sum_{n=1}^N q_n \sum_{k=1}^K [\nabla F(w^r)-\nabla F_n(w_n^{r,k})]||^2] \\
&\le \frac{K}{N} \sum_{n=1}^N \sum_{k=1}^K \mathbb{E}[||\nabla F(w^r)-\nabla F_n(w_n^{r,k})||^2] \\
&\le \frac{KL^2}{N}\sum_{n=1}^N \sum_{k=1}^K \mathbb{E}[||w^r-w_n^{r,k}||^2] \\
&\le 4\eta^2 L^2 G^2 K^4
 \end{aligned}
\vspace{-2mm}
\end{equation} 

When \( \eta < \frac{N}{PKL} \), we have: 
\begin{equation}
\small
\vspace{-1mm}
 \begin{aligned}
\mathbb{E}\![F(w^r)\!\!-\!\!F(w^{r-1]})] \!\!\le\!\! \frac{K\!L \!\eta^2\! \delta^2\!}{2N}\!\!+\!\!\frac{\!2\!N\!L^2\!G^2\!\eta^3\!K^3\!}{P}\!\!-\!\!\frac{\eta P K}{2 N}\mathbb{E}[||\nabla F(w^r)||^2]
 \end{aligned}
\vspace{-2mm}
\end{equation} \\
Rearrange the terms yields:
\begin{equation}
\small
\vspace{-1mm}
 \begin{aligned}
\!\mathbb{E}\textbf{}[||\!\nabla \!F(w^r)\!||^2\!] \!\le\! \frac{2N}{\eta P\!K}\! \mathbb{E}[F(w^{r-1})\!-\!F(w^r))]\!+\!\frac{\!L\eta\delta^2\!}{P}\!+\!\frac{\!4\!N^2\!L^2\!G^2\!\eta^2\!K^2\!}{P^2}
 \end{aligned}
\end{equation} \\
Telescoping from $r=1,...,R$, we can get the theorem 1.
\end{IEEEproof}

\subsection{Model drift with DGA in heterogeneous environments}
In the convergence upper bound of DGA (Theorem 2.3 in [9]), there is a variable \( D \) representing the number of delayed steps before global correction. In a homogeneous system, \( D \) is only related to the communication performance of devices, making it easy to control \( D \) within a small, manageable range. However, in a heterogeneous system, due to the significant differences in computation and communication performance across devices, \( D \) becomes uncontrollable. When the performance disparity among devices is substantial, \( D \) can become very large, rendering the upper bound given in Theorem 2.3 of [9] very loose, leading to failed convergence.
In contrast, in our FedEx framework, thanks to the proposed staleness ceiling, the convergence upper bound given in our Theorem 1 does not involve \( D \), but instead has a fixed upper bound. Therefore, FedEx ensures efficient operation even in heterogeneous systems.

\section{Analysis of FedEx's Algorithm Complexity}

FedEx primarily involves the computation of the utility function and overlapping trigger, which can be completed with an \textbf{O(1)} complexity. Moreover, both the utility computation and PS operations are executed on the resource-rich server. Since mobile devices only need to receive instructions from the server, FedEx does not introduce additional computational complexity to the mobile devices.

\section{Additional Experiments on Impacts of System Heterogeneity}
To further explore the performance of FedEx under different types of heterogeneity, we conducted additional experiments using five types of devices, ranked from highest to lowest performance: Xiaomi 12S, Xavier, Honor 70, Honor Play6T, and TX2. We considered six different heterogeneous device combinations, each consisting of 100 devices. In the first combination, 60 devices were Xiaomi 12S, while the remaining four types had 10 devices each. The second combination had 60 Xavier devices, with the rest evenly distributed. Similarly, the third, fourth, and fifth combinations had 60 Honor 70, Honor Play6T, and TX2 devices, respectively, with the remaining devices balanced across the other types. The sixth combination included an equal distribution of 20 devices for each type. We compared the overall latency required to reach a target accuracy of 44\%, and the results in Appendix Table II showed that FedEx consistently achieved the shortest latency under all configurations, demonstrating its effectiveness in heterogeneous system environments.

\begin{figure}
\renewcommand{\figurename}{Appendix Figure}
\centering
\includegraphics[width=0.45\textwidth]{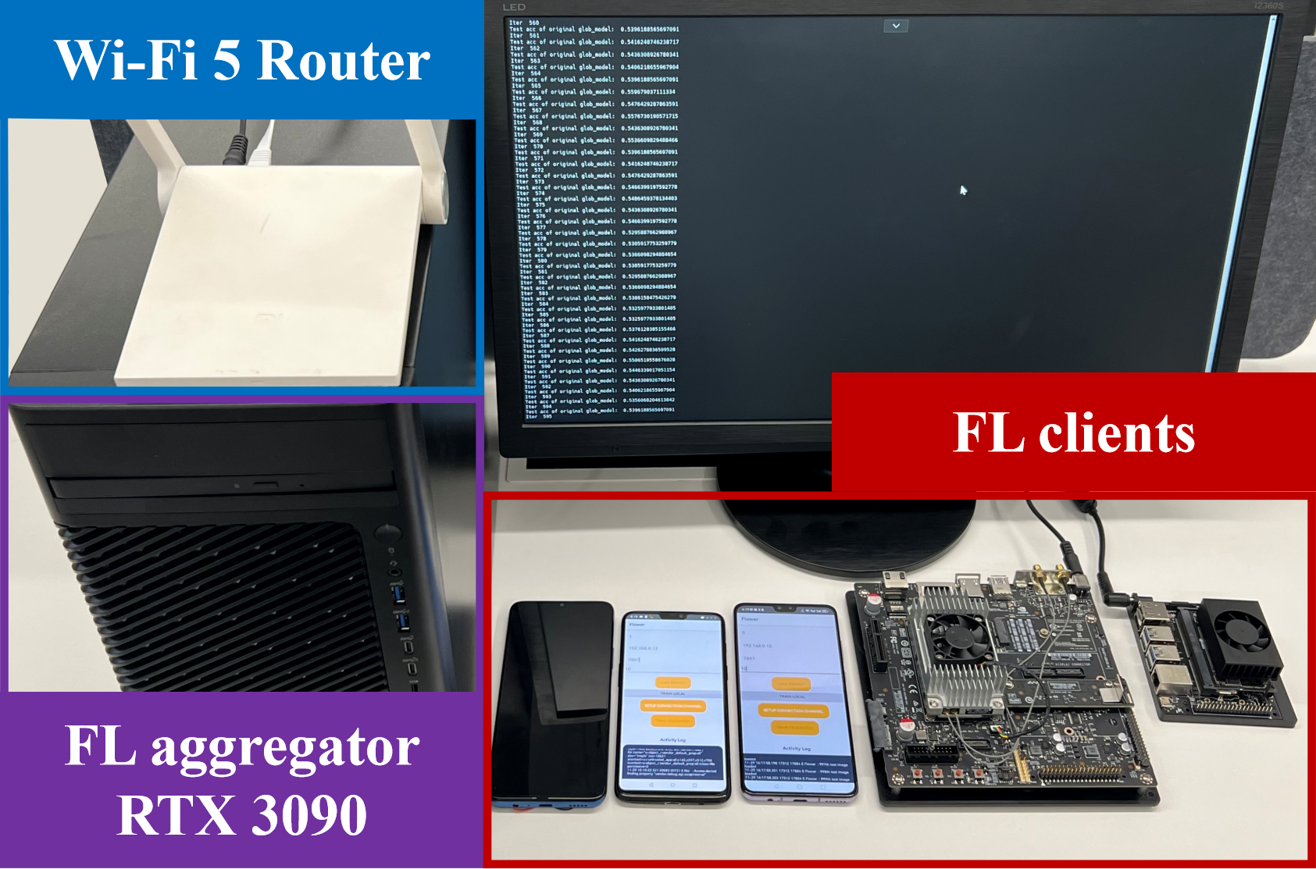}
\caption{FedEx Testbed.}
\label{fig:FedExArch}
\end{figure}

\section{FedEx Implementation Details}
The FedEx system has been implemented on a testbed, featuring an FL aggregator and a set of heterogeneous mobile devices, as shown in Fig.~\ref{fig:FedExArch}. The FL server utilizes a NVIDIA RTX 3090. 
On the FL client side, we have incorporated 5 types of mobile devices: 
(1) Xiaomi 12S smartphone with Qualcomm Snapdragon 8+ Gen 1 CPU, Adreno 740 GPU, and 8GB RAM; 
(2) Honor 70 smartphone with Qualcomm Snapdragon 778G Plus CPU, Qualcomm Adreno 642L GPU, and 8GB RAM; 
(3) Honor Play 6T smartphone with MediaTed Dimensity 700 CPU, Mali-G57 MC2 GPU, and 8GB RAM; 
(4) NVIDIA Jetson Xavier NX with 6-core NVIDIA Carmel ARM CPU, 384-core NVIDIA Volta GPU, and 8GB RAM; 
(5) NVIDIA Jetson TX2 with Dual-core NVIDIA Denver CPU, 256-core NVIDIA Pascal GPU, and 4GB RAM. 
The FedEx system encompasses a total of 100 mobile devices, with 20 devices from each type mentioned above. 
For each training round, 20 devices are selected.

For communication between FL clients and the FL server, 
a sophisticated wireless transmission scenario has been implemented, incorporating a blend of Wi-Fi 5, 4G, and 5G networks.
Specifically, NVIDIA Jetson Xavier NX and NVIDIA Jetson TX2 are connected to the FL server via Wi-Fi 5, utilizing the WebSocket communication protocol. 
The Xiaomi 12S smartphone is connected to the FL server via 5G, following the 5G NR standard. 
The Honor 70 smartphone and Honor Play 6T smartphone are connected to the FL server via 4G, following the 4G LTE standard. 
We further consider two types of transmission rates (high and low) under 4G or 5G settings.
Each device can be configured to one of these rates, corresponding to indoor and outdoor scenarios.
As for a Wi-Fi transmission environment, 
the nmcli and wondershaper command lines tools are employed for reporting network status and controlling various wireless transmission rates. 

\section{Related Work}
\textbf{FL with participant selection.} To mitigate the impact of stragglers in FL training, researchers have devoted significant efforts to developing device selection algorithms~\cite{appendix1, appendix2, appendix3}. Some strategies focus on selecting mobile devices showing significant training improvements, leveraging metrics like gradient divergence~\cite{appendix4}, local loss values, gradient norms~\cite{appendix5}, and the disparity between local and global models~\cite{appendix6, appendix7} to gauge their importance. Although these importance-aware scheduling policies have the advantage of communication round minimization, the heterogeneity of participating devices in their communication and computational capacity can still prolong convergence time due to the existence of straggler devices. To achieve FL speedup, strategies involve selecting devices with good instantaneous channel conditions to minimize communication latency~\cite{appendix8, appendix9}. Alternatively, certain approaches opt for high-speed devices to ensure local training completion within predefined timeframes~\cite{appendix10}. Due to the inconsistency between data distribution and resource distribution, it is possible that the subset of devices with superior resources may be unable to accommodate all data categories. Lai et al.'s approach~\cite{appendix11}, known as Oort, extends the selection criteria by considering both statistical and system utility, mitigating the negative effects of data and system heterogeneity on FL training. There are also endeavors aimed at enhancing model aggregation efficiency by enabling the central server to assess the computing capabilities of devices and select participants with well-performing models~\cite{appendix12}. However, these methods adhere to traditional FL training protocols, where devices are required to wait to upload local updates after completing their local computations. In other words, they have not fully tapped into the potential of utilizing such the waiting time to perform additional local computations that could significantly contribute to the convergence of training.

\textbf{FL with overlapping computation and communications.} Overlapping learning is an innovative strategy to accelerate FL training by allowing concurrent and continuous local computation during model uploading, mitigating the issue of training slowdowns caused by extended transmission delay. Approaches to achieve this include identifying optimal gradient transfer orders~\cite{appendix13}, leveraging stale model weights~\cite{appendix14}, and introducing delayed model updates~\cite{appendix15, appendix16}. As a notable overlapping scheme, DGA~\cite{appendix16} addresses challenges of model staleness in homogeneous FL environments by replacing local gradients from previous iterations with stale averaged gradients, ensuring more consistent and up-to-date model synchronization across devices. Advanced methods like AOCC-FL~\cite{appendix17} focus on aligning local models with the global model through calibrated compensation to minimize the gap between them. FedCR introduces contrastive learning during overlap training to reduce overlap staleness and accelerate FL training~\cite{appendix18}. An approach in~\cite{appendix19} proposes asynchronous global model aggregation by parallelizing server processes. However, existing methods often inadequately tackle issues like model drift and straggler problems introduced by data and device heterogeneity.




\end{document}